\documentclass[11pt]{article}

\usepackage[
a4paper,
margin=1in
]{geometry}

\usepackage{amsmath,amssymb,amsthm}
\usepackage[nolists,nomarkers]{endfloat}
\usepackage{bm}
\usepackage[authoryear, round]{natbib}
\usepackage{setspace}
\usepackage{enumitem}

\setlist[enumerate]{
	label=\arabic*.,
	leftmargin=*,
	align=left
}

\usepackage[figuresright]{rotating}

\usepackage{titling}
\usepackage[
colorlinks=true,
linkcolor=blue,
citecolor=blue,
urlcolor=black
]{hyperref}

\newtheorem{theorem}{Theorem}
\newtheorem{assumption}{Assumption}

\pretitle{\centering\Large\bfseries}
\posttitle{\par\vspace{2em}}

\preauthor{\centering\small}
\postauthor{\par\vspace{-0.5em}}

\predate{\begin{center}\small}

\makeatletter
\renewcommand\thanks[1]{\protect\footnotetext{#1}}
\makeatother

\makeatletter
\long\def\@makefntext#1{%
	\noindent\@makefnmark~#1
}
\makeatother

\usepackage[T1]{fontenc}
\usepackage{graphicx}
\usepackage{booktabs}
\usepackage{array}
\usepackage{float}
\usepackage{ragged2e}
\usepackage{caption}
\providecommand{\figalttext}[1]{}

\title{Modified Egger Intercept Tests for Detecting Horizontal Pleiotropy in Two-Sample Summary-Data Mendelian Randomization}
\author{
\textbf{Yilei Ma$^{1,2*}$}\thanks{*Yilei Ma, Youpeng Su and Xin Liu contributed equally to this work.},
\textbf{Youpeng Su$^{1*}$},
\textbf{Xin Liu$^{1*}$},
\textbf{Xuanye Cui$^{1}$},\\
\textbf{Siqi Xu$^{3}$},
\textbf{Ping Yin$^{1}$},
\textbf{Peng Wang$^{1\dagger}$}\thanks{$^{\dagger}$Corresponding author: Peng Wang. Email: \href{mailto:pengwang_stat@hust.edu.cn}{pengwang\_stat@hust.edu.cn}}\\[1em]
\parbox{0.96\textwidth}{\centering $^{1}$Department of Epidemiology and Biostatistics, School of Public Health, Tongji Medical College, Huazhong University of Science and Technology, Wuhan, China}\\[4pt]
\parbox{0.96\textwidth}{\centering $^{2}$Tongji Hospital, Tongji Medical College, Huazhong University of Science and Technology, Wuhan, China}\\[4pt]
\parbox{0.96\textwidth}{\centering $^{3}$Clinical Research Centre, Guangzhou First People's Hospital, Guangzhou Medical University, Guangzhou, China}\\[4pt]
}
\date{}

\begin{document}
\maketitle

\vspace{-2em}

\begin{abstract}
The Egger intercept (EI) test is a widely used tool to detect horizontal pleiotropy in two-sample summary-data Mendelian randomization. A significant test result suggests that either the average pleiotropic effect differs from zero (i.e., directional pleiotropy) or the InSIDE (Instrument Strength Independent of Direct Effect) assumption is violated (i.e., correlated pleiotropy) or both. As such, the EI test provides a diagnostic tool for assessing violations of the instrumental variable assumptions, as a nonzero intercept may indicate bias in the commonly used inverse-variance weighted (IVW) estimator. However, the EI test may exhibit inflated type I error rates due to estimation bias in Egger regression arising from measurement error and winner’s curse. In this article, we propose a modified EI (MEI) test based on a bias-corrected EI estimator under the null hypothesis of no directional or correlated pleiotropy, leveraging the recently developed rerandomized IVW estimator. We then establish the asymptotic properties of the proposed test under appropriate conditions. Like the EI test, we find that the power of the MEI test is also affected by SNP orientation. To improve the robustness of power, we further develop strategies for selecting appropriate allele coding schemes. Both simulation and real-data studies demonstrate that the MEI tests outperform the EI test in terms of type I error control and power.

\end{abstract}

\section{Introduction \label{section intro}}
Mendelian randomization (MR), an application of instrumental variable (IV) analysis in biomedical and epidemiological research, has become a powerful approach for inferring causal relationships between exposures and health outcomes \citep{smith2004mendelian}. By leveraging single nucleotide polymorphisms (SNPs) as IVs, MR can mitigate unmeasured confounding bias in observational studies \citep{didelez2007mendelian}. As extensive genome-wide association study (GWAS) summary data are publicly available, two-sample MR has become increasingly popular \citep{burgess2015blueprint, boehm2022methodsreview}.
In such a design, the summary statistics for the exposure and outcome are gleaned from two independent GWASs without requiring individual-level data. Various statistical methods have been developed for two-sample MR \citep{sanderson2022mendelian}. Among them, the inverse-variance weighted (IVW) method remains one of the most frequently used approaches due to its simplicity and efficiency \citep{burgess2013ivw}.
Despite its popularity, the validity of the IVW method relies on the assumption that all the included SNPs are valid IVs. That is, each SNP must satisfy the following three core assumptions:

\begin{enumerate}
	\item IV relevance: the SNP must be associated with the exposure;
	\item IV independence: the SNP must be independent of any confounders of the exposure-outcome relationship;
	\item Exclusion restriction: the SNP must affect the outcome only through the exposure.
\end{enumerate}

Without IV screening, the IV1 assumption is rarely satisfied for all SNPs because GWAS typically contain a subset of null and very weak IVs \citep{davies2015weakproblem}. The IVW method ignores the measurement error of the genetic association estimates for the exposure, and thus may suffer from measurement error bias (or weak instrument bias). To reduce such bias, practitioners usually only select SNPs that pass the genome-wide significance threshold ($P < 5\times10^{-8}$). However, the strategy of using the same exposure GWAS dataset for both IV selection and association estimation can additionally introduce winner’s curse bias \citep{gkatzionis2019contextualizing}. To address the above issues, some extensions of the IVW method have been proposed \citep{ye2021debiased, xu2023pena, ma2023breaking, su2024modified}. Of these, the rerandomized IVW (RIVW) estimator simultaneously corrects for measurement error bias and winner's curse bias without requiring any numerical optimization \citep{ma2023breaking}. Despite these advances, the extended methods built on the IVW framework still require that the IV2 and IV3 assumptions are both satisfied. However, violations of both assumptions are pervasive due to the widespread horizontal pleiotropy, as genetic variants may influence the outcome through multiple biological pathways \citep{solovieff2013pleichalleng, labrecque2018IVunderstanding, verbanck2018detection}. Specifically, the IV2 assumption may not hold due to residual population stratification in either the exposure or outcome GWAS, or associations between SNPs and heritable factors that confound the exposure--outcome relationship. In this scenario, the well-known InSIDE (Instrument Strength Independent of Direct Effect) assumption will break down, leading to correlated pleiotropy. Meanwhile, violation of the IV3 assumption can also happen when the SNP has a direct effect on the outcome. Nevertheless, this phenomenon does not affect the validity of the InSIDE assumption, and thus only introduces uncorrelated pleiotropy. Notably, although the IVW and its extended methods can be generalized to the scenario where the InSIDE assumption holds and the average pleiotropic effect is zero (i.e., balanced pleiotropy), this is only a special type of violation of the IV3 assumption.

To accommodate more general violations of the IV3 assumption, MR-Egger was proposed to account for directional pleiotropy while assuming that the InSIDE assumption holds.
MR-Egger involves three components: (1) a test for horizontal pleiotropy, (2) a test for the causal effect, and (3) an estimate for the causal effect \citep{burgess2017interpreting}. Although causal effect estimation and testing in MR-Egger require the InSIDE assumption \citep{slob2017note}, the test for horizontal pleiotropy does not rely on this assumption. This test assesses whether the intercept in Egger regression differs from zero and is thus referred to as the Egger intercept (EI) test (See more details in Section 3). Unlike other horizontal pleiotropy tests, such as Cochran's $Q$ test and the MR-PRESSO global test \citep{verbanck2018detection, bowden2019improving, wang2024powerful}, the EI test specifically targets directional or correlated pleiotropy, or both. Thus, the EI test offers a valuable diagnostic tool for assessing violations of the IV assumptions, with a nonzero EI indicating potential bias in the commonly used IVW estimator. However, estimation in MR-Egger is subject to the same sources of bias as the IVW estimator \citep{bowden2019meta}. Specifically, measurement error and winner's curse biases can attenuate the causal effect estimate toward zero while shifting the EI estimate away from zero, potentially resulting in inflated type I error rates in the EI test \citep{bowden2016suitability, ma2023breaking}.

In this article, we first derive a bias-corrected EI estimator under the null hypothesis of no directional or correlated pleiotropy, leveraging the RIVW estimator. Subsequently, we propose a modified EI (MEI) test by constructing a centered numerator-based statistic along with a closed-form estimator for its variance. We establish the asymptotic properties of the MEI test under appropriate conditions.
Like the EI test, we show that the power of the MEI test also depends on SNP orientation \citep{lin2021practical}.
To improve the robustness of power, we further develop strategies for selecting appropriate allele coding schemes.
Both simulation and real-data studies demonstrate that the MEI tests outperform the EI test in terms of type I error control and power.
Finally, we apply the proposed methods to examine horizontal pleiotropy in the relationships between complex metabolic traits and type 2 diabetes (T2D).

\section{Notation and assumptions\label{section notassump}}
\begin{figure}
	\begin{center}
		\centerline{\includegraphics[width=8.5cm]{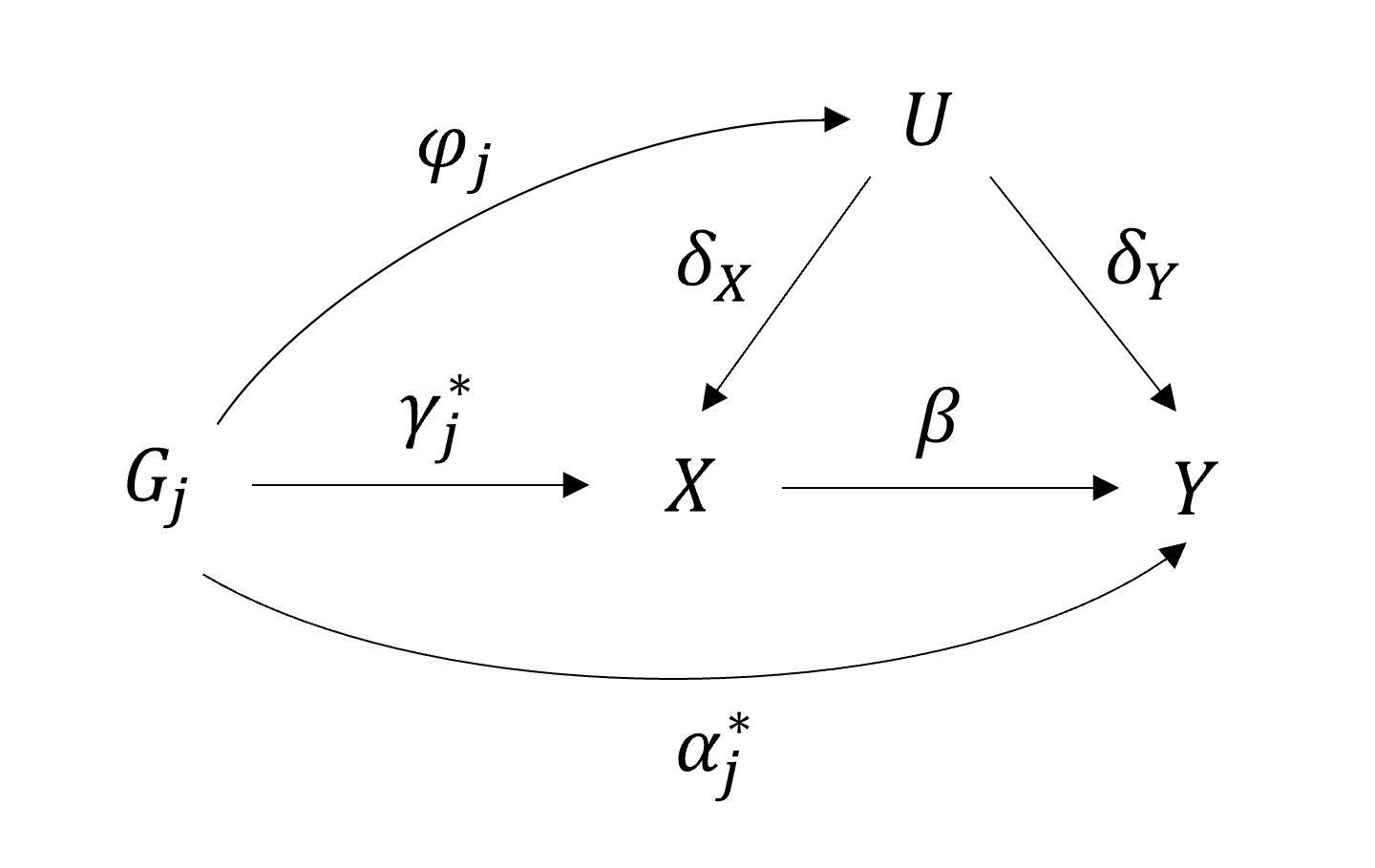}}
	\end{center}
	\caption{Causal model with SNP $G_j$, exposure $X$, outcome $Y$, and unmeasured confounder $U$.
		\label{fig:illu}}
\par\smallskip\noindent
\figalttext{Directed causal diagram linking a genetic variant to the exposure, outcome and an unmeasured confounder, with causal, direct and confounding pathways.}
\end{figure}

Suppose that after linkage disequilibrium (LD) clumping, we obtain $p$ independent SNPs $\{G_j \}_{j=1}^p$. For each SNP $G_j$, we assume that it is related to the exposure $X$, the outcome $Y$, and the unmeasured confounder $U$ through a linear causal model illustrated in Figure~\ref{fig:illu} \citep{bowden2017framework, xue2021constrain, xie2026robust}. Under this model, the effect of $G_j$ on $Y$, denoted by $\Gamma_j$, can be expressed as
\begin{eqnarray*}
	\Gamma_j = \beta \gamma_j + \alpha_j,
	\label{eq:sum}
\end{eqnarray*}
where $\beta$ denotes the causal effect of interest, $\gamma_j =\gamma_j^*+\delta_X \varphi_j$ represents the effect of $G_j$ on $X$, and $\alpha_j =\alpha_j^*+\delta_Y \varphi_j$ represents the pleiotropic effect, comprising both uncorrelated pleiotropy (the direct effect of $G_j$ on $Y$) and correlated pleiotropy (the effect of $G_j$ on $Y$ arising from confounding-induced associations).
Let $\bm{\gamma}=\{\gamma_j\}_{j=1}^p$ and $\bm{\alpha}=\{\alpha_j\}_{j=1}^p$.
Denote $\rho_{\gamma,\alpha}$ to be the correlation coefficient between $\bm{\gamma}$ and $\bm{\alpha}$, and $\mu_{\alpha}$ to be the mean of $\bm{\alpha}$, representing the average pleiotropic effect. Then, a non-zero $\rho_{\gamma, \alpha}$ indicates the presence of correlated pleiotropy (i.e., the InSIDE assumption is violated). Meanwhile, $\mu_{\alpha}$ provides another characterization of horizontal pleiotropy: balanced pleiotropy if $\mu_{\alpha}=0$ and directional pleiotropy if $\mu_{\alpha} \neq 0$. Notably, no horizontal pleiotropy can be viewed as a special case of balanced pleiotropy where the elements of $\bm{\alpha}$ are all zero.

In a two-sample MR design, the marginal regression coefficients of $X$ and $Y$ on $G_j$, denoted by $\hat{\gamma}_j$ and $\hat{\Gamma}_j$, respectively, and their corresponding standard errors, $\sigma_{X_j}$ and $\sigma_{Y_j}$, are obtained from the exposure and outcome GWASs. Throughout this paper, we adopt the following assumptions, which have also been considered in previous studies \citep{zhao2020raps, ye2021debiased, xu2023pena, ma2023breaking, su2024modified}.

\begin{assumption}\label{assump:infinity}
	The sample sizes for the exposure and outcome GWASs diverge to infinity with the same order. The number of SNPs, $p$, also diverges to infinity.
\end{assumption}

\begin{assumption}\label{assump:norm}
	For any $j \neq j'$, the pairs $(\hat{\Gamma}_j, \hat{\gamma}_j)$ and $(\hat{\Gamma}_{j'}, \hat{\gamma}_{j'})$ are mutually independent. For each $j$,
	\begin{eqnarray*}
		\left[
		\begin{array}{c}
			{\hat{\Gamma}_j} \\[0pt]
			{\hat{\gamma}_j}
		\end{array}\right]\
		\sim
		N\left(
		\left[
		\begin{array}{c}
			\beta \gamma_j + \alpha_j \\[0pt]
			\gamma_j
		\end{array}\right],
		\left[
		\begin{matrix}
			{\sigma_{Y_j}^2} & 0 \\[0pt]
			0 & {\sigma_{X_j}^2}
		\end{matrix}\right]\
		\right),
	\end{eqnarray*}
	with known variances $\sigma_{Y_j}^2$ and $\sigma_{X_j}^2$. In addition, there exists $\nu \to 0$ such that $\{\sigma_{X_j}/\nu,\; \sigma_{Y_j}/\nu: 1 \leq j \leq p\}$ are bounded and bounded away from zero.
\end{assumption}

\begin{assumption} \label{assump:norm-ii}
	The pleiotropic effects, $\alpha_1,\ldots,\alpha_p$, are mutually independent and follow a distribution with variance $\tau_{\alpha}^2$ and bounded third moment. In addition, $\tau_{\alpha} / \nu$ is bounded.
\end{assumption}

Assumption~\ref{assump:infinity} is reasonable in light of modern GWAS, which typically involve hundreds of thousands of participants and retain thousands of SNPs following LD clumping. Such large sample sizes provide the justification for Assumption \ref{assump:norm}, which posits the normality of $\hat{\Gamma}_j$ and $\hat{\gamma}_j$ together with their known variances. Furthermore, LD clumping enforces adequate separation between SNPs in terms of genetic distance, and each variant explains only a small proportion of the variance in the exposure and the outcome. These establish reasonable independence among the $\hat{\Gamma}_j$s and $\hat{\gamma}_j$s \citep{zhao2020raps, ye2021debiased}. The common scale $\nu$ in Assumption~\ref{assump:norm} specifies the order of $\sigma_{X_j}$ and $\sigma_{Y_j}$ and ensures that their ratio is uniformly bounded away from zero and infinity. The independence between $\hat{\Gamma}_j$ and $\hat{\gamma}_j$ relies on a two-sample MR design with non-overlapping samples. Assumption~\ref{assump:norm-ii} states that both the third moment and $\tau_\alpha/\nu$ are bounded, which is necessary for the subsequent asymptotic analysis. These conditions serve primarily as a technical convenience rather than a substantive biological assumption.

\section{Review of the EI test}
MR-Egger is implemented by fitting a weighted linear regression of the SNP--outcome association estimates on the SNP--exposure association estimates, with weights $\sigma_{Y_j}^{-2}$:
\begin{eqnarray*}
	\hat{\Gamma}_j = \beta \hat{\gamma}_j + \mu_{\alpha} + \varepsilon_j, \quad
	\varepsilon_j \sim N(0,  \sigma_{Y_j}^2 \phi),
\end{eqnarray*}
where $\varepsilon_j$ denotes the residual term and $\phi$ is an overdispersion parameter.
Under the InSIDE assumption, we can obtain consistent estimates of $\beta$ and $\mu_{\alpha}$, respectively denoted by $\hat{\beta}_\mathrm{E}$ and $\hat{\mu}_{\alpha, \mathrm{E}}$, along with their empirical covariance matrix from the weighted linear regression. In MR-Egger, the EI test evaluates whether the intercept $\mu_{\alpha}$ is significantly different from zero. Its test statistic is
\begin{eqnarray*}
	Z_\mathrm{E}=\frac{\hat{\mu}_{\alpha, \mathrm{E}}}{\sqrt{\hat{\mathrm{Var}}(\hat{\mu}_{\alpha, \mathrm{E}})}},
\end{eqnarray*}
where $\hat{\mathrm{Var}}(\hat{\mu}_{\alpha, \mathrm{E}})$ is the empirical variance of $\hat{\mu}_{\alpha, \mathrm{E}}$. Notably, under the InSIDE assumption (i.e., $\rho_{\gamma,\alpha}=0$) and when $\mu_{\alpha}=0$, $Z_\mathrm{E}$ asymptotically follows the standard normal distribution. Therefore, we can formulate the null hypothesis of the EI test as
\begin{eqnarray*}
	H_0: \mu_{\alpha} = 0 \quad \mathrm{and} \quad \rho_{\gamma, \alpha} = 0,
\end{eqnarray*}
which includes the scenario of no horizontal pleiotropy as a special case. The null hypothesis $H_0$ is composed of two conditions. Rejection of $H_0$ suggests the presence of directional or correlated pleiotropy or both. Consequently, the alternative hypothesis of the EI test is
\begin{eqnarray*}
	H_1: \mu_{\alpha} \neq 0 \quad \mathrm{or} \quad \rho_{\gamma, \alpha} \neq 0.
\end{eqnarray*}
Unlike other horizontal pleiotropy tests, such as Cochran’s $Q$ test and the MR-PRESSO global test, the EI test has no power to detect balanced pleiotropy when the InSIDE assumption holds.
Thus, the EI test provides a useful diagnostic tool for assessing the suitability of the IVW-based methods for MR analysis. However, because estimation in MR-Egger does not account for measurement error and winner's curse, the EI test may exhibit inflated type I error rates. To overcome this limitation, we propose a MEI test based on a bias-corrected EI estimator derived under $H_0$, leveraging the RIVW estimator.

\section{Methods}
\label{SectionMethods}
\subsection{The MEI test}
To motivate the proposed MEI test, we first express $\hat{\mu}_{\alpha, \mathrm{E}}$ in terms of $\hat{\beta}_\mathrm{E}$:
\begin{align}
	\hat{\mu}_{\alpha, \mathrm{E}} = \frac{\sum_{j=1}^p \sigma_{Y_j}^{-2} (\hat{\Gamma}_j - \hat{\beta}_\mathrm{E} \hat{\gamma}_j)}{\sum_{j=1}^p \sigma_{Y_j}^{-2}}.
	\label{eq:meta}
\end{align}
Because $\hat{\beta}_\mathrm{E}$ is subject to measurement error bias and winner's curse bias, $\hat{\mu}_{\alpha, \mathrm{E}}$ will also be affected by these sources of bias. To address this issue, a natural approach is to replace $\hat{\beta}_{\mathrm{E}}$ in Equation (\ref{eq:meta}) with an estimator of the causal effect $\beta$ that corrects for these biases.
We choose the RIVW estimator, denoted by $\hat{\beta}_\mathrm{R}$, as the plug-in estimate (see Web Appendix A for details).
This choice is motivated by two considerations.
First, the RIVW estimator is a consistent estimator for $\beta$ under $H_0$ and corrects for both measurement error and winner's curse biases.
Second, the RIVW estimator typically employs a less stringent threshold to select IVs (e.g., $P<5 \times 10^{-5}$), thus allowing the inclusion of more SNPs with moderate effect to improve the estimation precision.
Accordingly, we obtain the following modified EI estimator under $H_0$:
\begin{align*}
	\hat{\mu}_{\alpha,\mathrm{R}}
	=& \frac{\sum_{j \in \mathcal{S}_\lambda} \sigma_{Y_j}^{-2} (\hat{\Gamma}_j - \hat{\beta}_{\mathrm{R}} \hat{\gamma}_{j, \mathrm{RB}} )}
	{\sum_{j \in \mathcal{S}_\lambda} \sigma_{Y_j}^{-2}} \\
	=& \frac{\sum_{j \in \mathcal{S}_\lambda} \sigma_{Y_j}^{-2} \hat{\Gamma}_j
		\sum_{j \in \mathcal{S}_\lambda} \sigma_{Y_j}^{-2} \left( \hat{\gamma}_{j,\mathrm{RB}}^2 - \hat{\sigma}_{X_j,\mathrm{RB}}^2 \right)
		-
		\sum_{j \in \mathcal{S}_\lambda} \sigma_{Y_j}^{-2}  \hat{\gamma}_{j,\mathrm{RB}} \hat{\Gamma}_j
		\sum_{j \in \mathcal{S}_\lambda} \sigma_{Y_j}^{-2} \hat{\gamma}_{j,\mathrm{RB}}}
	{\sum_{j \in \mathcal{S}_\lambda} \sigma_{Y_j}^{-2}
		\sum_{j \in \mathcal{S}_\lambda} \sigma_{Y_j}^{-2} \left( \hat{\gamma}_{j,\mathrm{RB}}^2 - \hat{\sigma}_{X_j,\mathrm{RB}}^2 \right)},
	\label{eq:MEI}
\end{align*}
where the second equality follows by substituting the expression of $\hat{\beta}_\mathrm{R}$ into the first expression.
Here, $\mathcal{S}_\lambda$ denotes the set of SNPs selected under the screening rule adopted by the RIVW method, defined as  $\mathcal{S}_\lambda = \left\{j: |\hat{\gamma}_j/\sigma_{X_j} + Z_j| - \lambda > 0, j=1, \ldots ,p \right\}$, where $\lambda$ is a prespecified selection threshold and $Z_j \sim N(0,\eta^2)$ represents the corresponding pseudo SNP-exposure association effect.
Note that $\eta$ is a preset tuning parameter that is assumed to be bounded away from zero and infinity. Throughout this article, we fix $\eta$ to be 0.5, as it leads to satisfactory performance for the RIVW estimator \citep{ma2023breaking}.
The estimator $\hat{\gamma}_{j,\mathrm{RB}}$, derived by applying Rao–Blackwellization, achieves minimum variance for all unbiased estimators for $\gamma_j$ conditional on the selection event.
Its conditional variance is denoted by $\sigma_{X_j,\mathrm{RB}}^2$, and
$\hat{\sigma}_{X_j,\mathrm{RB}}^2$ is an unbiased estimator for $\sigma_{X_j,\mathrm{RB}}^2$ given the selection event (see Web Appendix A for details).

Because the modified estimator $\hat{\mu}_{\alpha,\mathrm{R}}$ is free of both measurement error and winner's curse biases, it is expected to be close to zero under $H_0$. This implies that its numerator
\begin{align*}
	\hat{\Lambda}_\mathrm{R} = \sum_{j \in \mathcal{S}_\lambda} \sigma_{Y_j}^{-2} \hat{\Gamma}_j
	\sum_{j \in \mathcal{S}_\lambda} \sigma_{Y_j}^{-2} \left( \hat{\gamma}_{j,\mathrm{RB}}^2 - \hat{\sigma}_{X_j,\mathrm{RB}}^2 \right)
	-
	\sum_{j \in \mathcal{S}_\lambda} \sigma_{Y_j}^{-2}  \hat{\gamma}_{j,\mathrm{RB}} \hat{\Gamma}_j
	\sum_{j \in \mathcal{S}_\lambda} \sigma_{Y_j}^{-2} \hat{\gamma}_{j,\mathrm{RB}}
\end{align*}
should be near to zero either. For mathematical convenience, we may consider using $\hat{\Lambda}_\mathrm{R}$ instead of  $\hat{\mu}_{\alpha,\mathrm{R}}$  to construct the test statistic. However, simple algebra reveals that the conditional expectation of $\hat{\Lambda}_\mathrm{R}$ under $H_0$ is  $E(\hat{\Lambda}_\mathrm{R}|\mathcal{S}_\lambda)= -\sum_{j \in \mathcal{S}_\lambda} \sigma_{Y_j}^{-4} \sigma_{X_j,\mathrm{RB}}^2 \Gamma_j$, which is nonzero. Hence, we make adjustment of $\hat{\Lambda}_\mathrm{R}$ and derive the following estimator:
\begin{align*}
	\hat{\Lambda}_{\mathrm{R, C}} = \hat{\Lambda}_{\mathrm{R}} + \sum_{j \in \mathcal{S}_\lambda} \sigma_{Y_j}^{-4} \hat{\sigma}_{X_j,\mathrm{RB}}^2 \hat{\Gamma}_j.
\end{align*}
Under $H_0$, the conditional expectation of $\hat{\Lambda}_{\mathrm{R, C}}$ becomes exactly zero. We therefore construct the MEI test using this adjusted statistic.
We next turn to its variance estimation. For this purpose, we make the following decomposition:
\begin{align*}
	\hat{\Lambda}_{\mathrm{R, C}} \approx \sum_{j \in \mathcal{S}_\lambda} {\sigma_{Y_j}^{-2} u_j},
\end{align*}
\begin{align*}
	u_j =
	\left(\hat{\Gamma}_j - \beta \hat{\gamma}_{j,\mathrm{RB}}\right)
	\sum_{l \in \mathcal{S}_\lambda} \sigma_{Y_l}^{-2} \gamma_l^2
	-
	\left[\hat{\gamma}_{j,\mathrm{RB}} \hat{\Gamma}_j - \beta \left(\hat{\gamma}_{j,\mathrm{RB}}^2 - \hat{\sigma}_{X_j,\mathrm{RB}}^2 \right)\right]
	\sum_{l \in \mathcal{S}_\lambda} \sigma_{Y_l}^{-2} \gamma_l.
\end{align*}
Web Appendix B provides a detailed theoretical justification for this
decomposition under appropriate conditions,
establishing
$\sum_{j\in\mathcal S_\lambda}\sigma_{Y_j}^{-2}u_j$
as the leading term of $\hat{\Lambda}_{\mathrm{R,C}}$.
Conditional on $\mathcal{S}_\lambda$, the $u_j$s are mutually independent with mean zero and different variances under $H_0$.
Consequently, the conditional variance of the leading term is
\begin{align*}
	V_{\mathrm{R, C}}
    = \operatorname{Var}\left(\sum_{j \in \mathcal{S}_\lambda} \sigma_{Y_j}^{-2}u_j \mid \mathcal{S}_\lambda\right)
	= \sum_{j \in \mathcal{S}_\lambda} \sigma_{Y_j}^{-4}\operatorname{Var}(u_j \mid \mathcal{S}_\lambda),
\end{align*}
and a moment-based estimator for $V_{\mathrm{R, C}}$ can be obtained using the sum-of-squares form $\sum_{j \in \mathcal{S}_\lambda} \sigma_{Y_j}^{-4} u_j^2$.
Since each $u_j$ involves the same set of unknown parameters, we replace them by their corresponding estimators and derive the following variance estimator:
\begin{align*}
	\hat{V}_{\mathrm{R, C}} &= \sum_{j \in \mathcal{S}_\lambda} {\sigma_{Y_j}^{-4} \hat{u}_j^2},\\
	\hat{u}_j &=
	\left(\hat{\Gamma}_j - \hat{\beta}_{\mathrm{R}} \hat{\gamma}_{j,\mathrm{RB}}\right)
	\sum_{k \in \mathcal{S}_\lambda} \sigma_{Y_k}^{-2}
	\left(\hat{\gamma}_{k,\mathrm{RB}}^2 - \hat{\sigma}_{X_k,\mathrm{RB}}^2 \right)\\
	&\quad -
	\left[
	\hat{\gamma}_{j,\mathrm{RB}} \hat{\Gamma}_j
	- \hat{\beta}_{\mathrm{R}}
	\left(\hat{\gamma}_{j,\mathrm{RB}}^2 - \hat{\sigma}_{X_j,\mathrm{RB}}^2 \right)
	\right]
	\sum_{k \in \mathcal{S}_\lambda} \sigma_{Y_k}^{-2} \hat{\gamma}_{k,\mathrm{RB}}
	.
\end{align*}
We can then construct the MEI test statistic as
\begin{align*}
	Z_{\mathrm{ME}} = \frac{\hat{\Lambda}_{\mathrm{R, C}}}{\sqrt{\hat{V}_{\mathrm{R, C}}}}.
\end{align*}
To establish its asymptotic normality, we introduce the following additional assumption.

\begin{assumption} \label{assump:nodom}

	The true instrument effects satisfy
	\[
	\frac{\max_{j \in \mathcal{S}_\lambda}(\gamma_j-\bar{\gamma}_\lambda)^2}
	{\sum_{j \in \mathcal{S}_\lambda}(\gamma_j-\bar{\gamma}_\lambda)^2}
	\xrightarrow{p} 0,
	\] where $\bar{\gamma}_\lambda	= {p_\lambda}^{-1} \sum_{j \in \mathcal{S}_\lambda}\gamma_j
	$ denotes the average true SNP--exposure effect among the $p_\lambda$ selected SNPs.
\end{assumption}

Assumption~\ref{assump:nodom} rules out individual instruments that dominate the variation in the true SNP--exposure effects around their mean after selection. This condition ensures that no single instrument contributes disproportionately to the centered variation, facilitating the establishment of asymptotic normality for the intercept statistic.
Define the average instrument strength as $\kappa_\lambda = {p_\lambda}^{-1} \sum_{j \in \mathcal{S}_\lambda} \sigma_{X_j}^{-2} \gamma_j^2$. The following theorem establishes the asymptotic normality of the MEI test statistic.

\begin{samepage}
\begin{theorem}\label{theo}
	Suppose that Assumptions \ref{assump:infinity}--\ref{assump:nodom} hold, $p_\lambda \xrightarrow{p} \infty$ and $\kappa_\lambda\sqrt{p_\lambda} \xrightarrow{p} \infty$. Under $H_0$, $Z_{\mathrm{ME}}$ is asymptotically standard normal, i.e.,
	\[
	Z_{\mathrm{ME}} \xrightarrow{d} N(0, 1).
	\]
\end{theorem}
\end{samepage}
The proof of Theorem~\ref{theo} is provided in Web Appendix C. Accordingly, the asymptotic $P$-value can be calculated using the standard normal distribution.

 Notably, \citet{ma2023breaking} established
	asymptotic properties for the RIVW estimator under	$\lambda\to\infty$ and
	$\kappa_\lambda/\lambda^2\xrightarrow{p}\infty$, along with
	other regularity conditions.
	Building on this framework, we allow a fixed $\lambda$ and instead require $\kappa_\lambda\sqrt{p_\lambda}\xrightarrow{p}\infty$.
	These conditions relax the requirements on the selection threshold and average instrument strength, allowing scenarios with many weak instruments \citep{ye2021debiased}.
	Under the stated assumptions, we establish the consistency of
	$\hat{\beta}_{\mathrm{R}}$ under the relaxed conditions in
	Web Appendix A and use this result to prove the asymptotic normality of
	$Z_{\mathrm{ME}}$.
	Further details on the asymptotic properties under these
	weaker conditions are provided by \citet{liu2026brivw}, who extend the RIVW framework to account for sample structure.

\subsection{Choice of allele coding strategies and the statistical power}
Like the EI test, the MEI test depends on SNP orientation \citep{lin2021practical}. Specifically, changing the reference allele for an included SNP may alter the resulting MEI test statistic. This dependence suggests that allele coding schemes may also affect the test power. To illustrate this, we derive the conditional expectation of $\hat{\Lambda}_{\mathrm{R, C}}$ under $H_1$:
\begin{eqnarray}
	E(\hat{\Lambda}_{\mathrm{R, C}} \mid \mathcal{S}_\lambda)
	\propto
	\mu_{\alpha|\mathcal{S}_\lambda} \tau_{\gamma|\mathcal{S}_\lambda}^2
	- \rho_{\gamma,\alpha|\mathcal{S}_\lambda} \mu_{\gamma|\mathcal{S}_\lambda} \tau_{\alpha|\mathcal{S}_\lambda} \tau_{\gamma|\mathcal{S}_\lambda},
	\label{eq:expect}
\end{eqnarray}
where $\mu_{\alpha|\mathcal{S}_\lambda}$ and $\mu_{\gamma|\mathcal{S}_\lambda}$ denote the means of $\{\alpha_j\}_{j \in \mathrm{S}_\lambda}$ and $\{\gamma_j\}_{j \in \mathrm{S}_\lambda}$, respectively, $\tau_{\alpha|\mathcal{S}_\lambda}$ and $\tau_{\gamma|\mathcal{S}_\lambda}$ denote their standard deviations, and $\rho_{\gamma,\alpha|\mathcal{S}_\lambda}$ is their correlation coefficient (see Web Appendix D for details).
As can be seen, the orientations of SNPs affect the definitions of all the parameters in Equation (\ref{eq:expect}) under $H_1$, and thus may further impact the effect size and power. Therefore, we must carefully specify the coding strategy for the included SNPs before calculating the test statistic.

Identifying the optimal allele coding strategy that maximizes the power of the MEI test is challenging. For practical purposes, we consider two specific allele coding schemes. The first is major-allele-referenced coding, which uses the major allele as the reference allele for all included SNPs. Because GWAS summary data typically provide the allele frequency information, such coding strategy can be readily implemented in MR studies. Notably, for a highly polygenic trait, under major-allele-referenced coding, it is reasonable to model SNP effects as arising from a distribution symmetric about zero, provided that minor alleles are equally likely to increase or decrease the trait.
If the exposure satisfies this condition, symmetry of the SNP--exposure effects and the selection event implies $\mu_{\gamma|\mathcal{S}_\lambda}=0$, so the second term in Equation~(\ref{eq:expect}) vanishes. Consequently, the power of the MEI test under major-allele-referenced coding is driven by directional pleiotropy, whereas correlated pleiotropy contributes no signal. Notably, directional pleiotropy alone does not bias the RIVW estimator in this scenario. Thus, the corresponding MEI test cannot serve as a tool to assess the suitability of the RIVW estimator.
To address this limitation, we additionally consider all-positive coding, which uses the exposure-decreasing allele as the reference allele for all included SNPs. In practice, selected SNPs typically have relatively strong SNP--exposure associations, thus we can accurately identify the effect-decreasing allele for a specific SNP. This coding strategy is also commonly used in MR-Egger \citep{burgess2017interpreting}.
It ensures that all $\hat{\gamma}_{j}$s are positive and shifts $\mu_{\gamma|\mathcal{S}_\lambda}$ away from zero
to the greatest extent, thereby enhancing the
	power to detect correlated pleiotropy. Moreover, in the absence of correlated pleiotropy, directional pleiotropic effects cancel out under all-positive coding assuming the SNP-exposure effects are symmetric about zero. Consequently, the MEI test under all-positive coding can be interpreted as a test specifically targeting correlated pleiotropy under this symmetry assumption. Therefore, we recommend all-positive coding when the exposure is a highly polygenic trait.

However, if the exposure is not highly polygenic, the assumption that SNP--exposure effects follow a symmetric distribution with mean zero may no longer be reasonable. In such scenarios, the MEI tests under major-allele-referenced coding and all-positive coding can no longer be interpreted as tests specifically targeting directional and correlated pleiotropy, respectively. Nevertheless, the two tests may offer complementary advantages, with one potentially being more powerful in detecting directional pleiotropy and the other in detecting correlated pleiotropy. We therefore combine them using a maximum $Z$-score approach.
Specifically, let $Z_{\mathrm{ME}}^{\mathrm{M}}$ and $Z_{\mathrm{ME}}^{\mathrm{P}}$ respectively denote the $Z$-scores obtained under the major-allele-referenced coding and all-positive coding and define
\begin{eqnarray*}
	Z_\mathrm{ME}^\mathrm{C} = \max\left(|Z_{\mathrm{ME}}^{\mathrm{M}}|, |Z_{\mathrm{ME}}^{\mathrm{P}}|\right)
\end{eqnarray*}
as the combined test statistic.
Under $H_0$, $Z_{\mathrm{ME}}^{\mathrm M}$ and $Z_{\mathrm{ME}}^{\mathrm P}$ are asymptotically jointly bivariate normal with standard normal marginals. We estimate their correlation coefficient $\rho_{\mathrm{MP}}$ by
\[
\hat{\rho}_{\mathrm{MP}}
=
\frac{
	\sum_{j\in\mathcal S_\lambda}
	\sigma_{Y_j}^{-4}\hat u_j^{\mathrm M}\hat u_j^{\mathrm P}
}{
	\sqrt{
		\left[
		\sum_{j\in\mathcal S_\lambda}
		\sigma_{Y_j}^{-4}(\hat u_j^{\mathrm M})^2
		\right]
		\left[
		\sum_{j\in\mathcal S_\lambda}
		\sigma_{Y_j}^{-4}(\hat u_j^{\mathrm P})^2
		\right]
	}
},
\]
where $\hat u_j^{\mathrm M}$ and $\hat u_j^{\mathrm P}$ denote the versions of $\hat u_j$ obtained under the major-allele-referenced and all-positive coding, respectively.
Given an observed value of $Z_\mathrm{ME}^\mathrm{C}$, denoted by $z$, the corresponding $P$-value is calculated by
\begin{eqnarray*}
	P(Z_\mathrm{ME}^\mathrm{C} > z) = 1 - P(-z \leq Z_{\mathrm{ME}}^{\mathrm{M}} \le z, -z \leq Z_{\mathrm{ME}}^{\mathrm{P}} \le z),
\end{eqnarray*}
which is a bivariate normal integral. Its calculation is available in the commonly used software (e.g., mvtnorm package in $\bm{\mathrm{R}}$).
By integrating the test statistics derived from two different coding schemes, the combined test aims to maintain high power for detecting both directional and correlated pleiotropy, as will be shown in the simulation study.

\section{Simulations}
\label{SectionSimulations}
\subsection{Simulation settings}
\noindent To assess the type I error rate and power of the proposed MEI tests ($Z_\mathrm{ME}^\mathrm{M}$, $Z_\mathrm{ME}^\mathrm{P}$, and $Z_\mathrm{ME}^\mathrm{C}$), an extensive simulation study is conducted for comparison with the EI test ($Z_\mathrm{E}$) and the modified Cochran's $Q$ test ($Q$) with exact weighting proposed by \cite{bowden2019improving}. Note that the null hypothesis for the $Q$ test is no horizontal pleiotropy, which is different from those for the other tests.
We independently generate summary-level data for $200,000$ candidate SNPs from $\hat{\gamma}_j\sim N(\gamma_j,\sigma_{X_j}^2)$ and $\hat{\Gamma}_j\sim N(\beta\gamma_j+\alpha_j,\sigma_{Y_j}^2)$.
The parameters $(\gamma_j,\alpha_j)$ are assumed to jointly follow the mixture distribution
\begin{eqnarray*}
	\begin{pmatrix}\gamma_j\\ \alpha_j\end{pmatrix}
	&\sim&
	\pi_1q
	\begin{pmatrix}N(\mu_x,\tau_x^2)\\ N(0,\tau_y^2)\end{pmatrix}
	+\pi_1(1-q)
	\begin{pmatrix}N(\mu_x,\tau_x^2)\\ \delta_0\end{pmatrix}
	+\pi_2q
	\begin{pmatrix}\delta_0\\ N(0,\tau_y^2)\end{pmatrix}\\
	&&+\pi_2(1-q)
	\begin{pmatrix}\delta_0\\ \delta_0\end{pmatrix}
	+\pi_3r
	\begin{pmatrix}N(\mu_x,\tau_x^2)\\ N(\mu_y,\tau_y^2)\end{pmatrix}
	+\pi_3(1-r)
	\begin{pmatrix}N(\mu_x,\tau_x^2)+A_j\\ N(0,\tau_y^2)+A_j\end{pmatrix},
\end{eqnarray*}
where $\delta_0$ is the Dirac measure centred at zero, and $A_j\sim N(0,s^2)$.
As such, correlated pleiotropy is introduced through $A_j$, with $s^2$ controlling the magnitude of the shared effect.
The mixture proportions satisfy $\pi_1+\pi_2+\pi_3=1$. Under $H_0$, we set $\pi_3=0$, which results in a four-component mixture model. The first two terms correspond to relevant IVs, and the third and fourth terms represent SNPs with no exposure effects. The parameter $q$ controls the proportion of SNPs exhibiting balanced pleiotropy within each group.
Under $H_1$, we allow $\pi_3>0$ to introduce directional and/or correlated pleiotropy, with $r$ controlling the proportion of directional pleiotropic IVs. Specifically, we consider three scenarios: (i) directional pleiotropy only ($r=1$), (ii) correlated pleiotropy only ($r=0$), and (iii) both types of pleiotropy ($0<r<1$).

 We set $\mu_x=0$ to represent a balance between positive and negative SNP--exposure effects, as expected for highly polygenic traits, and $\mu_x=0.005$ to allow deviations from this symmetry.
	Type I error rates are evaluated under $\pi_3=0$, with $\pi_1\in\{5\%,10\%,20\%\}$, under both scenarios without balanced pleiotropy ($q=0$) and with balanced pleiotropy ($q=10\%$).
	Power is evaluated under $\pi_3 \in \{0.5\%,1\%,2\%\}$ across the three scenarios described above, with $r$ taking values 1, 0, and $\{20\%,50\%,80\%\}$, respectively.
	The same settings for $\pi_1$ and $q$ are considered, with power results reported for $\pi_1=10\%$.
	Directional pleiotropy is introduced by setting $\mu_y=-0.015$, whereas correlated pleiotropy is introduced by setting $s^2=2\times10^{-5}$.
 The exposure and outcome GWAS sample sizes are set to be equal, with $n = 250,000$ or $500,000$. The true causal effect is fixed at $\beta=0.5$.
Minor allele frequencies $f_j$ are independently generated from a uniform distribution over $(0.1,0.5)$, and the exposure and outcome variances are both set to 1. The summary-data variances are specified as
$
	\sigma_{X_j}^{2}
	=[{1-2f_j(1-f_j)\gamma_j^2}] / [{2nf_j(1-f_j)}]
$
and
$
	\sigma_{Y_j}^{2}
	=[{1-2f_j(1-f_j)(\beta\gamma_j+\alpha_j)^2}] / [{2nf_j(1-f_j)}].
$
The SNP--exposure association variance and the pleiotropic variance are fixed at \(\tau_x^2=2\times10^{-5}\) and \(\tau_y^2=4\times10^{-5}\), respectively.
These parameter settings yield expected exposure and outcome heritabilities ranging from below \(5\%\) (low) to approximately \(70\%\) (high).

Without loss of generality, we assume that all simulated data are generated under the major-allele-referenced coding scheme.
For the EI test, we adopt all-positive coding, which is recommended and implemented in several popular MR software packages \citep{yavorska2017Rpackage,hemani2018platform}.
To mitigate measurement error bias, IVs for the EI test are selected using the conventional threshold of $\lambda=5.45$, corresponding to a significance level of $5\times10^{-8}$.
The same IV selection threshold is used for the $Q$ test, whose statistic is invariant to SNP orientation.
In contrast, the three MEI tests utilize the RIVW estimator to account for potential biases, thereby enabling the inclusion of SNPs with weaker effect sizes.
Accordingly, a less stringent threshold of $\lambda=4.06$ is adopted at the significance level $5\times10^{-5}$ for the MEI tests.
Finally, we perform $10,000$ simulation replicates and set the significance level at 0.05 for all tests.

\subsection{Simulation results}

Table~\ref{tab:type1error} presents the estimated type I error rates of various tests under $\mu_x=0$.
From the table, we find that all three MEI tests maintain accurate type I error rates in all the situations.
	In contrast, the EI test exhibits inflated type I error rates due to measurement error and winner's curse, with inflation becoming more pronounced as the GWAS sample size ($n$) or the proportion of relevant IVs ($\pi_1$) increases.
	For example, when $q=0$ and $\pi_1=20\%$, the type I error rate for the EI test increases from 7.82\% to 14.78\% as $n$ increases from 250,000 to 500,000.
	Similarly, the $Q$ test fails to control the type I error rates in scenarios without horizontal pleiotropy $(q=0)$.  Web Table S1 in the Supplementary Materials reports the estimated type I error rates of various tests under $\mu_x=0.005$. As shown, all three MEI tests continue to maintain accurate type I error rates, whereas those of the EI and $Q$ tests generally exhibit larger deviations from the nominal level.

\begin{figure}
	\begin{center}
		\centerline{\includegraphics[width=\linewidth]{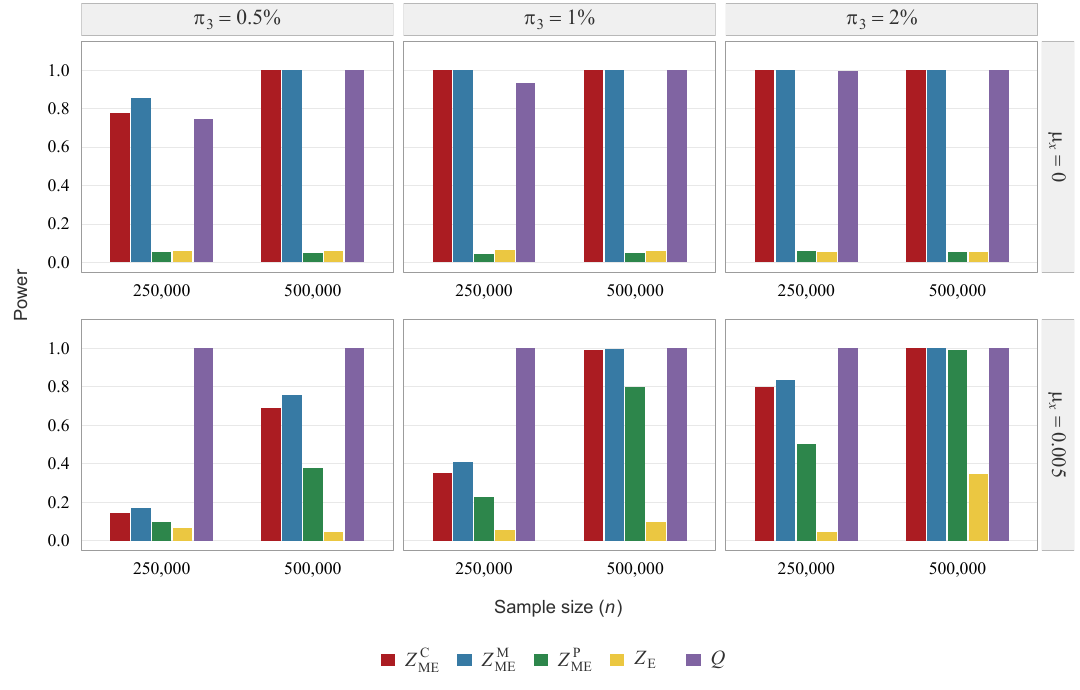}}
	\end{center}
	\caption{Estimated power of various tests to detect directional pleiotropy ($r=1$) based on $10,000$ replicates. There are no balanced pleiotropic IVs ($q = 0$).}
	\label{fig:PowerDirect}
	\par\smallskip\noindent
	\figalttext{Grouped bar charts comparing the power of the five tests to detect directional pleiotropy, for proportions of directional pleiotropic IVs of 0.5, 1 and 2 percent, GWAS sample sizes of 250,000 and 500,000, and mean SNP--exposure effects among relevant IVs of 0 and 0.005.}
\end{figure}

Figure \ref{fig:PowerDirect} plots the estimated power of various tests to detect directional pleiotropy when $q=0$.
From the figure, we find that $Z_{\mathrm{ME}}^{\mathrm{M}}$ has the best performance in power among the three MEI tests.
The powers of $Z_{\mathrm{ME}}^{\mathrm{C}}$ are slightly lower than those of $Z_{\mathrm{ME}}^{\mathrm{M}}$, but they are overall comparable.
Note that $Z_{\mathrm{ME}}^{\mathrm{P}}$ exhibits much lower power than $Z_{\mathrm{ME}}^{\mathrm{M}}$. When $\mu_x=0$, $Z_{\mathrm{ME}}^{\mathrm{P}}$ has no power to detect directional pleiotropy because the directional pleiotropic effect cancels out under all-positive coding.
When $\mu_x=0.005$, $Z_{\mathrm{E}}$ is less powerful than $Z_{\mathrm{ME}}^{\mathrm{P}}$, possibly because its more stringent IV selection threshold reduces the number of included instruments. Notably, the $Q$ test maintains high power across all scenarios. However, because the $Q$ and MEI tests target different null hypotheses, their performance cannot be compared solely on the basis of power.

\begin{figure}
	\begin{center}
		\centerline{\includegraphics[width=\linewidth]{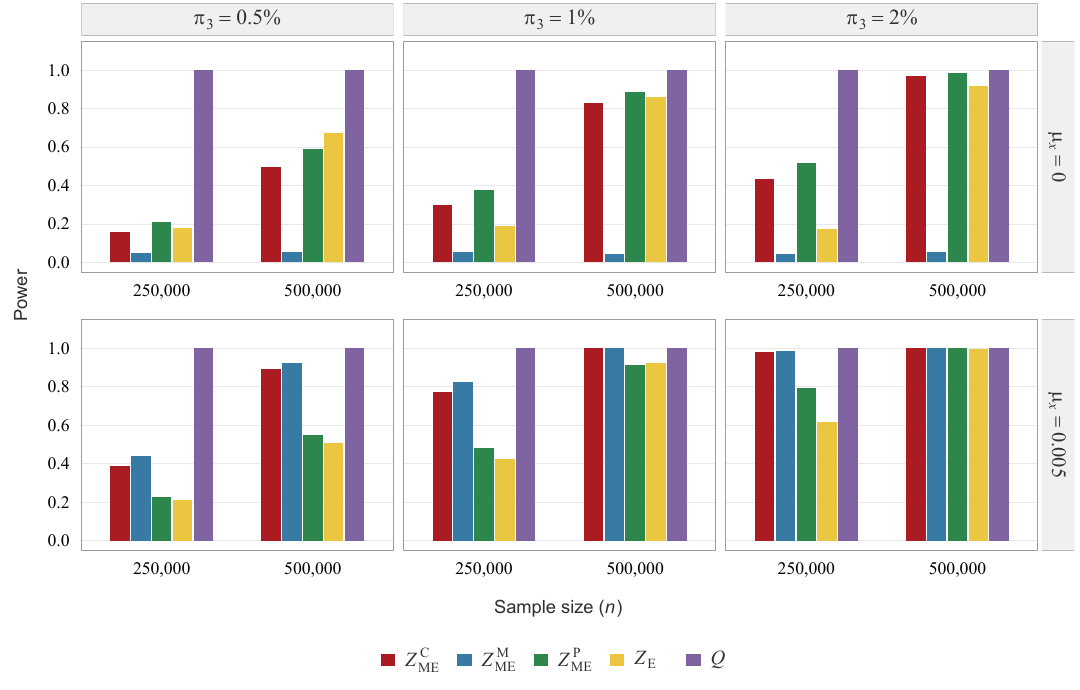}}
	\end{center}
	\caption{Estimated power of various tests to detect correlated pleiotropy ($r=0$) based on $10,000$ replicates. There are no balanced pleiotropic IVs ($q = 0$).}
	\label{fig:PowerCorr}
	\par\smallskip\noindent
	\figalttext{Grouped bar charts comparing the power of the five tests to detect correlated pleiotropy, for proportions of correlated pleiotropic IVs of 0.5, 1 and 2 percent, GWAS sample sizes of 250,000 and 500,000, and mean SNP--exposure effects among relevant IVs of 0 and 0.005.}
\end{figure}

Figure \ref{fig:PowerCorr} displays the estimated power of various tests to detect correlated pleiotropy when $q=0$.
Unlike the power patterns observed in Figure \ref{fig:PowerDirect}, $Z_{\mathrm{ME}}^{\mathrm{P}}$ achieves the highest power among all three MEI tests when $\mu_x=0$, whereas $Z_{\mathrm{ME}}^{\mathrm{M}}$ has no power beyond the nominal significance level, consistent with our theoretical results.
As $\mu_x$ increases to 0.005, the power of $Z_{\mathrm{ME}}^{\mathrm{M}}$ grows rapidly and even exceeds that of $Z_{\mathrm{ME}}^{\mathrm{P}}$.
As expected, the power of $Z_{\mathrm{ME}}^{\mathrm{C}}$ remains close to the higher power of $Z_{\mathrm{ME}}^{\mathrm{M}}$ and $Z_{\mathrm{ME}}^{\mathrm{P}}$.
Although $Z_{\mathrm{E}}$ and $Z_{\mathrm{ME}}^{\mathrm{P}}$ achieve similar power when $n=500,000$, the latter generally exhibits higher power at the smaller sample size.
In both figures, when $n$ increases from $250,000$ to $500,000$, and $\pi_3$ grows from 0.5\% to 2\%, the powers of the MEI tests tend to become larger.
The $Q$ test also maintains high power to detect correlated pleiotropy.

\begin{figure}
	\begin{center}
		\centerline{\includegraphics[width=\linewidth]{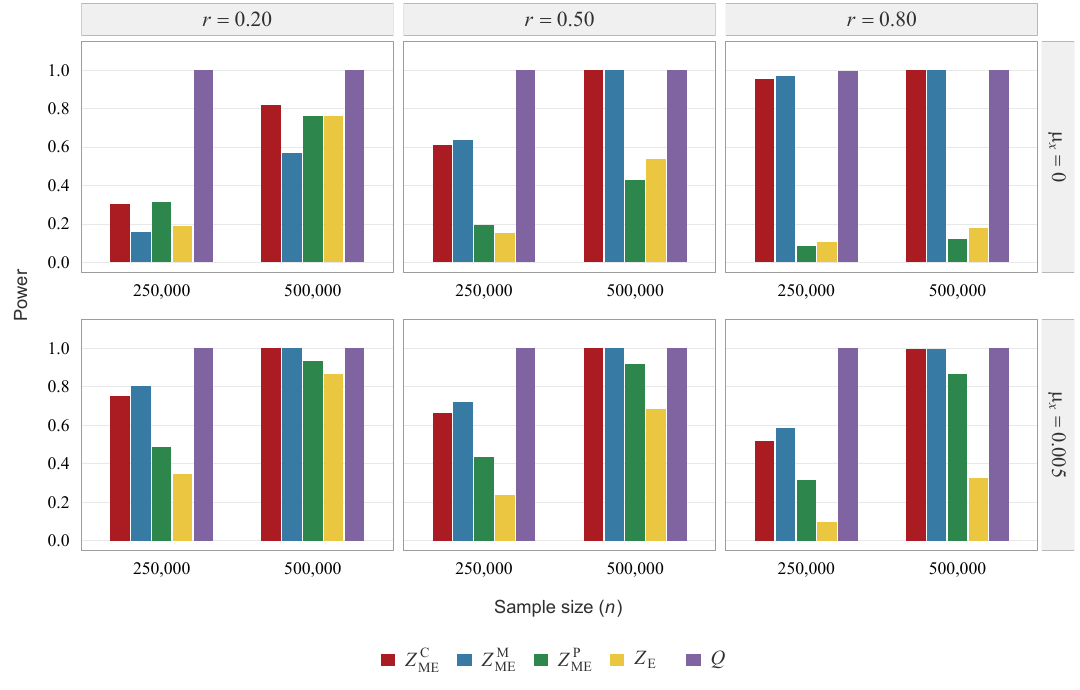}}
	\end{center}
	\caption{Estimated power of various tests in the presence of directional and correlated pleiotropy based on $10,000$ replicates. The total proportion of directional and correlated pleiotropic IVs is fixed at $\pi_3 = 1\%$, and there are no balanced pleiotropic IVs ($q = 0$).}
	\label{fig:PowerDC}
\par\smallskip\noindent
\figalttext{Grouped bar charts comparing the power of the five tests as the proportion of directional pleiotropic IVs among all IVs with directional or correlated pleiotropy increases from 20 to 50 to 80 percent, for GWAS sample sizes of 250,000 and 500,000, and mean SNP--exposure effects among relevant IVs of 0 and 0.005.}
\end{figure}

Figure \ref{fig:PowerDC} presents the estimated power when both directional and correlated pleiotropy are present, with $q=0$ and $\pi_3=1\%$. When $\mu_x=0$, the power of $Z_{\mathrm{ME}}^{\mathrm{M}}$ increases while that of $Z_{\mathrm{ME}}^{\mathrm{P}}$ decreases as $r$ varies from 20\% to 80\%.
When $\mu_x=0.005$, the powers of all three MEI tests decrease with increasing $r$ at $n=250,000$, but remain high at $n=500,000$.
The $Q$ test maintains power close to 100\% across these mixed pleiotropy settings.
Additional power results with $q=10\%$ are provided in Supplementary Materials (see Web Figures S1--S3). Compared with results in Figures \ref{fig:PowerDirect}–\ref{fig:PowerDC}, the three MEI tests and the EI test exhibit a slight reduction in power in the presence of balanced pleiotropic IVs, while the main conclusions remain unchanged.

\section{Real Data Applications}
T2D is a complex disease that imposes a substantial global health burden, affecting over 500 million adults worldwide and contributing to premature mortality \citep{kalyani2025T2DReview}.
Circulating metabolites are involved in key biological pathways underlying T2D development, including insulin resistance and glucose homeostasis \citep{accili2025T2Dreview}.
Consistent with these mechanisms, epidemiological evidence suggests that metabolic traits may contribute to T2D risk, with further support from MR studies \citep{liu2017MRStudy, ahola2019cohorts}.
However, given the complexity of both metabolic traits and T2D, causal inference using MR requires careful assessment of potential pleiotropic effects that may violate the core IV assumptions.
In this section, we assess the presence of directional and/or correlated pleiotropy in MR analyses of metabolic traits and T2D.
GWAS summary statistics for 249 metabolic measures, including lipoprotein lipids, fatty acids, and small molecules, are obtained from the UK Biobank, involving $115,082$ participants \citep{richardson2022characterising}.
The T2D GWAS statistics are obtained from the FinnGen study (release R12), including $82,878$ cases and $403,489$ controls \citep{Kurki2023FinnGen}.
The exposure and outcome datasets are derived from non-overlapping samples of similar ancestry.
Linkage disequilibrium clumping is performed to remove correlated IVs ($r^2 < 0.001$ within a 10-Mb window).

Given the polygenic architecture of most metabolic traits, we apply the Wilcoxon signed-rank test to assess whether the SNP--exposure effects after IV selection are symmetrically distributed around zero under major-allele-referenced coding. This approach is motivated by the relatively small measurement errors in the selected SNP--exposure association estimates compared with the corresponding effect sizes. After Rao--Blackwell correction, the mean of the selected SNP--exposure effect estimates, averaged across 249 metabolic traits, is $2.2\times10^{-5}$. The Wilcoxon signed-rank test yields $P<0.05$ for only 9 of the 249 traits (3.6\%). These findings are consistent with the assumption that the selected SNP--exposure effects after IV selection are approximately symmetric around zero for most metabolic traits.

We then apply the three MEI tests, the EI test, and the $Q$ test using the same IV selection threshold as in the simulations. Statistical significance is assessed at both the nominal level ($P < 0.05$) and the Bonferroni-corrected threshold ($P < 0.05/249 \approx 2 \times 10^{-4}$).
The results are summarized in Table~\ref{tab:pleiotropy_detection2}.
In the analysis of 249 metabolic trait--T2D pairs, $Z_{\mathrm{ME}}^{\mathrm P}$ identifies widespread evidence of pleiotropy, with 177 pairs (71.1\%) being statistically significant and 106 pairs (42.6\%) remaining significant after Bonferroni correction.
$Z_{\mathrm{ME}}^{\mathrm C}$ identifies slightly fewer significant pairs than $Z_{\mathrm{ME}}^{\mathrm P}$,
 whereas $Z_{\mathrm{ME}}^{\mathrm M}$ detects very few significant pairs.
This observation is not surprising because T2D is also a highly polygenic trait and the directional pleiotropic effects are expected to be small, with
$
\mu_{\alpha\mid\mathcal{S}_\lambda}
=
\mu_{\Gamma\mid\mathcal{S}_\lambda}
-\beta\mu_{\gamma\mid\mathcal{S}_\lambda}
\approx 0,
$
where $\mu_{\Gamma\mid\mathcal{S}_\lambda}$ denotes the mean
SNP--outcome effect among the selected SNPs.
These findings suggest that correlated pleiotropy may be a major concern in MR analyses of metabolic traits and T2D.
The EI test detects substantially fewer significant pairs, none of which remain significant after Bonferroni correction.
Its lower detection rate may be partly explained by the smaller number of selected IVs, averaging 49 compared with 318 for the MEI tests.
In addition, the $Q$ test identifies heterogeneity in 247 pairs (99.2\%) after Bonferroni correction.

To further assess the prevalence of pleiotropy among putative causal associations, we apply the RIVW method and identify 127 pairs at the nominal significance level, of which 35 remain significant after Bonferroni correction.
Among the 127 pairs with evidence of causal associations, 77 (60.6\%) also yield significant results from $Z_{\mathrm{ME}}^{\mathrm P}$ after Bonferroni correction.
When the analysis is restricted to the 35 pairs with Bonferroni-significant causal associations, this proportion increases to 82.9\%. The same proportion is observed for $Z_{\mathrm{ME}}^{\mathrm C}$ in this subset, retaining all associations detected by $Z_{\mathrm{ME}}^{\mathrm P}$. These findings underscore the importance of carefully assessing pleiotropy when interpreting MR-based causal effect estimates.

\section{Discussion}

Although the EI test is widely used to detect horizontal pleiotropy in two-sample MR, its performance can be adversely affected by biased parameter estimation in Egger regression arising from measurement error and winner’s curse, resulting in inflated type I error rates.
In this study, we propose a MEI test that improves type I error control by leveraging the RIVW estimator to construct a bias-corrected EI estimator under the null hypothesis.
Such bias correction also allows the MEI test to incorporate a larger set of IVs under a less stringent selection threshold, thereby improving its power.
Because the power of the MEI test depends on SNP orientation, we develop strategies for selecting appropriate allele coding schemes. Our findings suggest that all-positive coding is preferable when the exposure is a highly polygenic trait. Otherwise, we recommend a combined test based on the MEI test statistics obtained under major-allele-referenced coding and all-positive coding. Both simulation and real-data studies demonstrate that the MEI tests outperform the EI test in terms of type I error control and power.
In practice, the MEI test can serve as a useful diagnostic tool for assessing the suitability of IVW-based methods. When a statistically significant result is obtained, causal effect estimation methods that are robust to both directional and correlated pleiotropy, such as CARE \citep{xie2026robust}, are preferred for reliable MR analysis.

Several limitations warrant consideration. First, the MEI tests may have limited power in certain settings. As indicated by Equation~(\ref{eq:expect}), the signals arising from directional and correlated pleiotropy can partially cancel each other when $\mu_{\alpha|\mathcal{S}_\lambda}$ and $\rho_{\gamma,\alpha|\mathcal{S}_\lambda}\mu_{\gamma|\mathcal{S}_\lambda}$ have the same sign. The EI test is subject to a similar limitation. In the presence of such cancellation, larger exposure and outcome GWAS sample sizes may be needed to achieve adequate power.
Second, the present work focuses on addressing type I error inflation in the EI test arising from measurement error and winner’s curse, whereas the development of a corresponding causal effect estimator is beyond its scope. The latter is investigated in a separate ongoing study on the rerandomized Egger estimator, which is developed within a different theoretical framework \citep{su2026regger}.
Finally, as in many previous studies, we assume that the exposure and outcome GWAS samples are completely disjoint. However, applied MR analyses often use large consortium-based GWAS datasets, leading to sample overlap. In such cases, the SNP--exposure and SNP--outcome association estimates are no longer independent, thereby violating Assumption \ref{assump:norm}. Consequently, the proposed MEI tests may not maintain their nominal type I error rates in the presence of sample overlap. Addressing this issue is an important direction for future research.

\section*{Supplementary material}
Web Appendices and Figures referenced in Sections~\ref{SectionMethods} and~\ref{SectionSimulations} are provided in the Supplementary Materials accompanying this manuscript.
The R package for the MEI tests, including source code and example data, is publicly available at \url{https://github.com/biostacreator/mr.MEITests}.

\section*{Data availability}
The data supporting the findings of this paper are openly available for metabolic traits at \url{https://www.ebi.ac.uk/gwas/studies/GCST90092803} and for type 2 diabetes at \url{https://r12.finngen.fi/}.

\section*{Acknowledgements}
We acknowledge the UK Biobank and FinnGen consortia for making GWAS summary statistics publicly available.

\section*{Funding}
This work was supported by the National Natural Science Foundation of China [grant number 82504525]; and the Hubei Provincial Natural Science Foundation of China [grant number 2024AFB027].

\section*{Conflict of interest}
The authors declare no conflicts of interest.


\begin{table}
	\centering
	\setstretch{1.5}
	\caption{Estimated type I error rates (\%) of various methods at the nominal significance level of 0.05 under $\mu_x=0$ based on $10,000$ replicates.}
	\label{tab:type1error}
	\begin{tabular*}{\textwidth}{@{\extracolsep{\fill}}cccccccc@{}}
		\toprule
		$n$ & $q$ & $\pi_1$ & $Z_{\mathrm{ME}}^{\mathrm{C}}$ & $Z_{\mathrm{ME}}^{\mathrm{M}}$ & $Z_{\mathrm{ME}}^{\mathrm{P}}$ & $Z_{\mathrm{E}}$ & $Q^*$ \\
		\midrule
		250,000 & 0\% & 5\% & 4.98 & 4.72 & 5.35 & 5.59 & 7.14 \\
		& & 10\% & 5.32 & 5.12 & 5.32 & 6.68 & 8.08 \\
		& & 20\% & 4.77 & 4.89 & 4.84 & 7.82 & 9.67 \\

		& 10\% & 5\% & 5.38 & 5.17 & 5.62 & 6.59 & 31.39 \\
		& & 10\% & 5.45 & 5.09 & 5.51 & 6.39 & 45.57 \\
		& & 20\% & 5.05 & 5.04 & 4.88 & 7.24 & 65.19 \\

		500,000 & 0\% & 5\% & 5.00 & 4.54 & 5.00 & 7.52 & 9.46 \\
		& & 10\% & 5.21 & 5.14 & 4.99 & 9.95 & 11.63 \\
		& & 20\% & 5.02 & 5.11 & 5.09 & 14.78 & 15.44 \\

		& 10\% & 5\% & 4.99 & 5.05 & 5.26 & 6.78 & 97.05 \\
		& & 10\% & 4.90 & 4.94 & 5.10 & 7.91 & 99.90 \\
		& & 20\% & 5.31 & 5.62 & 5.03 & 10.85 & 100.00 \\
		\bottomrule
	\end{tabular*}
	\par\vspace{0.35\baselineskip}
	\parbox{\textwidth}{%
		\footnotesize\justifying\noindent
		*For \(Q\), the results under \(q=10\%\) represent its empirical power to detect balanced pleiotropy under the InSIDE assumption.
		\par
	}
\end{table}

\begin{table}
	\centering
	\setstretch{1.5}
	\caption{Application of various methods to detect directional and/or correlated pleiotropy in MR analysis. \# IVs: Counts of selected IVs averaged across all pairs.}
	\label{tab:pleiotropy_detection2}
	\begin{tabular*}{\textwidth}{@{}p{0.15\textwidth}@{}>{\centering\arraybackslash}p{0.17\textwidth}@{}>{\centering\arraybackslash}p{0.30\textwidth}@{}>{\centering\arraybackslash}p{0.38\textwidth}@{}}
		\toprule
		& &
		\multicolumn{2}{c}{\textbf{Counts of pleiotropic pairs (\%)}} \\
		\cmidrule(lr){3-4}
		& & \textbf{At 0.05} & \textbf{At Bonferroni} \\[-0.6em]
		\textbf{Method} & \textbf{\# IVs} &
		\textbf{significance level} & \textbf{corrected threshold} \\

		\midrule
		\multicolumn{4}{l}{\emph{249 pairs}} \\
		\hspace{0.5em}$Z_{\mathrm{ME}}^{\mathrm{C}}$ & 318 & 164 (65.9\%) & 95 (38.2\%) \\
		\hspace{0.5em}$Z_{\mathrm{ME}}^{\mathrm{M}}$ & & 15 (6.0\%) & 0 \\
		\hspace{0.5em}$Z_{\mathrm{ME}}^{\mathrm{P}}$ & & 177 (71.1\%) & 106 (42.6\%) \\
		\hspace{0.5em}$Z_{\mathrm{E}}$ & 49 & 12 (4.8\%) & 0 \\
		\hspace{0.5em}$Q$ &  & 248 (99.6\%) & 247 (99.2\%) \\

		\multicolumn{4}{l}{\emph{127 causal pairs identified by the RIVW estimator at the 0.05 significance level}} \\
		\hspace{0.5em}$Z_{\mathrm{ME}}^{\mathrm{C}}$ & 346 & 112 (88.2\%) & 71 (55.9\%) \\
		\hspace{0.5em}$Z_{\mathrm{ME}}^{\mathrm{M}}$ & & 8 (6.3\%) & 0 \\
		\hspace{0.5em}$Z_{\mathrm{ME}}^{\mathrm{P}}$ & & 115 (90.6\%) & 77 (60.6\%) \\
		\hspace{0.5em}$Z_{\mathrm{E}}$ & 59 & 10 (7.9\%) & 0 \\
		\hspace{0.5em}$Q$ &  & 127 (100.0\%) & 127 (100.0\%) \\

		\multicolumn{4}{l}{\emph{35 causal pairs identified by the RIVW estimator at the Bonferroni-corrected threshold}} \\
		\hspace{0.5em}$Z_{\mathrm{ME}}^{\mathrm{C}}$ & 382 & 34 (97.1\%) & 29 (82.9\%) \\
		\hspace{0.5em}$Z_{\mathrm{ME}}^{\mathrm{M}}$ & & 2 (5.7\%) & 0 \\
		\hspace{0.5em}$Z_{\mathrm{ME}}^{\mathrm{P}}$ & & 34 (97.1\%) & 29 (82.9\%) \\
		\hspace{0.5em}$Z_{\mathrm{E}}$ & 73 & 0 & 0 \\
		\hspace{0.5em}$Q$ &  & 35 (100.0\%) & 35 (100.0\%) \\

		\bottomrule
	\end{tabular*}
\end{table}

\begin{thebibliography}{99}

\bibitem[\protect\citeauthoryear{Accili et~al.}{2025}]{accili2025T2Dreview}
Accili, D., Deng, Z. and Liu, Q. (2025).
Insulin resistance in type 2 diabetes mellitus.
{\it Nature Reviews Endocrinology} {\bfseries 21}, 413--426.

\bibitem[\protect\citeauthoryear{Ahola-Olli et~al.}{2019}]{ahola2019cohorts}
Ahola-Olli, A.V., Mustelin, L., Kalimeri, M., Kettunen, J., Jokelainen, J., Auvinen, J., Puukka, K., Havulinna, A.S., Lehtimäki, T., Kähönen, M., Juonala, M., Keinänen-Kiukaanniemi, S., Salomaa, V., Perola, M., Järvelin, M.R., et~al. (2019).
Circulating metabolites and the risk of type 2 diabetes: a prospective study of 11,896 young adults from four Finnish cohorts.
{\it Diabetologia} {\bfseries 62}, 2298--2309.

\bibitem[\protect\citeauthoryear{Boehm and Zhou}{2022}]{boehm2022methodsreview}
Boehm, F.J. and Zhou, X. (2022).
Statistical methods for Mendelian randomization in genome-wide association studies: a review.
{\it Computational and Structural Biotechnology Journal} {\bfseries 20}, 2338--2351.

\bibitem[\protect\citeauthoryear{Bowden et~al.}{2016}]{bowden2016suitability}
Bowden, J., Del Greco M, F., Minelli, C., Davey Smith, G., Sheehan, N.A., and Thompson, J.R. (2016).
Assessing the suitability of summary data for two-sample Mendelian randomization analyses using MR-Egger regression: the role of the $I^2$ statistic.
{\it International Journal of Epidemiology} {\bfseries {45}}, 1961--1974.

\bibitem[\protect\citeauthoryear{Bowden et~al.}{2017}]{bowden2017framework}
Bowden, J., Del Greco M, F., Minelli, C., Davey Smith, G., Sheehan, N. and Thompson, J. (2017).
A framework for the investigation of pleiotropy in two-sample summary data Mendelian randomization.
{\it Statistics in Medicine} {\bfseries 36}, 1783--1802.

\bibitem[\protect\citeauthoryear{Bowden et~al.}{2019}]{bowden2019improving}
Bowden, J., Del Greco M, F., Minelli, C., Zhao, Q., Lawlor, D.A., Sheehan, N.A., Thompson, J. and Davey Smith, G. (2019).
Improving the accuracy of two-sample summary-data Mendelian randomization: moving beyond the NOME assumption.
{\it International Journal of Epidemiology} {\bfseries 48}, 728--742.

\bibitem[\protect\citeauthoryear{Bowden and Holmes}{2019}]{bowden2019meta}
Bowden, J. and Holmes, M.V. (2019).
Meta-analysis and Mendelian randomization: a review.
{\it Research Synthesis Methods} {\bfseries 10}, 486--496.

\bibitem[\protect\citeauthoryear{Burgess, Butterworth and Thompson}{Burgess et~al.}{2013}]{burgess2013ivw}
Burgess, S., Butterworth, A. and Thompson, S.G. (2013).
Mendelian randomization analysis with multiple genetic variants using summarized data.
{\it Genetic Epidemiology} {\bfseries 37}, 658--665.

\bibitem[\protect\citeauthoryear{Burgess et~al.}{2015}]{burgess2015blueprint}
Burgess, S., Scott, R.A., Timpson, N.J., Davey Smith, G., Thompson, S.G. and EPIC-InterAct Consortium. (2015).
Using published data in Mendelian randomization: a blueprint for efficient identification of causal risk factors.
{\it European Journal of Epidemiology} {\bfseries 30}, 543--552.

\bibitem[\protect\citeauthoryear{Burgess and Thompson}{2017}]{burgess2017interpreting}
Burgess, S. and Thompson, S.G. (2017).
Interpreting findings from Mendelian randomization using the MR-Egger method.
{\it European Journal of Epidemiology} {\bfseries 32}, 377--389.

\bibitem[\protect\citeauthoryear{Davey Smith and Ebrahim}{2004}]{smith2004mendelian}
Davey Smith, G. and Ebrahim, S. (2004).
Mendelian randomization: prospects, potentials, and limitations.
{\it International Journal of Epidemiology} {\bfseries 33}, 30--42.

\bibitem[\protect\citeauthoryear{Davies et~al.}{2015}]{davies2015weakproblem}
Davies, N.M., von Hinke Kessler Scholder, S., Farbmacher, H., Burgess, S., Windmeijer, F. and Davey Smith, G. (2015).
The many weak instruments problem and Mendelian randomization.
{\it Statistics in Medicine} {\bfseries 34}, 454--468.

\bibitem[\protect\citeauthoryear{Didelez and Sheehan}{2007}]{didelez2007mendelian}
Didelez, V. and Sheehan, N. (2007).
Mendelian randomization as an instrumental variable approach to causal inference.
{\it Statistical Methods in Medical Research} {\bfseries 16}, 309--330.

\bibitem[\protect\citeauthoryear{Gkatzionis and Burgess}{2019}]{gkatzionis2019contextualizing}
Gkatzionis, A. and Burgess, S. (2019).
Contextualizing selection bias in Mendelian randomization: how bad is it likely to be?
{\it International Journal of Epidemiology} {\bfseries 48}, 691--701.

\bibitem[\protect\citeauthoryear{Hemani et~al.}{2018}]{hemani2018platform}
Hemani, G., Zheng, J., Elsworth, B., Wade, K.H., Haberland, V., Baird, D., Laurin, C., Burgess, S., Bowden, J., Langdon, R., Tan, V.Y., Yarmolinsky, J., Shihab, H.A., Timpson, N.J., Evans, D.M., et~al. (2018).
The MR-Base platform supports systematic causal inference across the human phenome.
{\it eLife} {\bfseries 7}, e34408.

\bibitem[\protect\citeauthoryear{Kalyani et~al.}{2025}]{kalyani2025T2DReview}
Kalyani, R.R., Neumiller, J.J., Maruthur, N.M. and Wexler, D.J. (2025).
Diagnosis and treatment of type 2 diabetes in adults: a review.
{\it Journal of the American Medical Association} {\bfseries 334}, 984--1002.

\bibitem[\protect\citeauthoryear{Kurki et~al.}{2023}]{Kurki2023FinnGen}
Kurki, M.I., Karjalainen, J., Palta, P., Sipilä, T.P., Kristiansson, K., Donner, K.M., Reeve, M.P., Laivuori, H., Aavikko, M., Kaunisto, M.A., Loukola, A., Lahtela, E., Mattsson, H., Laiho, P., Della Briotta Parolo, P., et~al. (2023).
FinnGen provides genetic insights from a well-phenotyped isolated population.
{\it Nature} {\bfseries 613}, 508--518.

\bibitem[\protect\citeauthoryear{Labrecque and Swanson}{2018}]{labrecque2018IVunderstanding}
Labrecque, J. and Swanson, S.A. (2018).
Understanding the assumptions underlying instrumental variable analyses: a brief review of falsification strategies and related tools.
{\it Current Epidemiology Reports} {\bfseries 5}, 214--220.

\bibitem[\protect\citeauthoryear{Lin et~al.}{2022}]{lin2021practical}
Lin, Z., Pan, I. and Pan, W. (2022).
A practical problem with Egger regression in Mendelian randomization.
{\it PLoS Genetics} {\bfseries 18}, e1010166.

\bibitem[\protect\citeauthoryear{Liu et~al.}{2017}]{liu2017MRStudy}
Liu, J., van Klinken, J.B., Semiz, S., van Dijk, K.W., Verhoeven, A., Hankemeier, T., Harms, A.C., Sijbrands, E., Sheehan, N.A., van Duijn, C.M. and Demirkan, A. (2017).
A Mendelian randomization study of metabolite profiles, fasting glucose, and type 2 diabetes.
{\it Diabetes} {\bfseries 66}, 2915--2926.

\bibitem[\protect\citeauthoryear{Liu et~al.}{2026}]{liu2026brivw}
Liu, X., Su, Y., Zeng, C., Ma, Y., Xu, S., Yin, P. and Wang, P. (2026).
Simultaneously accounting for the winner's curse and sample structure in Mendelian randomization: bivariate rerandomized inverse variance weighted estimator.
{\it arXiv preprint}, arXiv:2603.06078.

\bibitem[\protect\citeauthoryear{Ma, Wang and Wu}{Ma et~al.}{2023}]{ma2023breaking}
Ma, X., Wang, J. and Wu, C. (2023).
Breaking the winner’s curse in Mendelian randomization: rerandomized inverse variance weighted estimator.
{\it The Annals of Statistics} {\bfseries 51}, 211--232.

\bibitem[\protect\citeauthoryear{Morrison et~al.}{2020}]{morrison2020mendelian}
Morrison, J., Knoblauch, N., Marcus, J.~H., Stephens, M. and He, X. (2020).
Mendelian randomization accounting for correlated and uncorrelated pleiotropic effects using genome-wide summary statistics.
{\it Nature Genetics}, {\bf 52}(7), 740--747.

\bibitem[\protect\citeauthoryear{Richardson et~al.}{2022}]{richardson2022characterising}
Richardson, T.G., Leyden, G.M., Wang, Q., Bell, J.A., Elsworth, B., Davey Smith, G. and Holmes, M.V. (2022).
Characterising metabolomic signatures of lipid-modifying therapies through drug target Mendelian randomisation.
{\it PLoS Biology} {\bfseries 20}, e3001547.

\bibitem[\protect\citeauthoryear{Sanderson et~al.}{2022}]{sanderson2022mendelian}
Sanderson, E., Glymour, M.M., Holmes, M.V., Kang, H., Morrison, J., Munafò, M.R., Palmer, T., Schooling, C.M., Wallace, C., Zhao, Q. and Davey Smith, G. (2022).
Mendelian randomization.
{\it Nature Reviews Methods Primers} {\bfseries 2}, 6.

\bibitem[\protect\citeauthoryear{Slob et~al.}{2017}]{slob2017note}
Slob, E.A.W., Groenen, P.J.F., Thurik, A.R. and Rietveld, C.A. (2017).
A note on the use of Egger regression in Mendelian randomization studies.
{\it International Journal of Epidemiology} {\bfseries 46}, 2094--2097.

\bibitem[\protect\citeauthoryear{Solovieff et~al.}{2013}]{solovieff2013pleichalleng}
Solovieff, N., Cotsapas, C., Lee, P.H., Purcell, S.M. and Smoller, J.W. (2013).
Pleiotropy in complex traits: challenges and strategies.
{\it Nature Reviews Genetics} {\bfseries 14}, 483--495.

\bibitem[\protect\citeauthoryear{Su et~al.}{2026}]{su2026regger}
Su, Y., Ma, Y., Yin, P. and Wang, P. (2026).
Rerandomized Egger estimator in two-sample summary-data Mendelian randomization.
{\it arXiv preprint}, arXiv:2507.09634v2.

\bibitem[\protect\citeauthoryear{Su et~al.}{2024}]{su2024modified}
Su, Y., Xu, S., Ma, Y., Yin, P., Hao, X., Zhou, J., Fung, W.K., Jiang, H. and Wang, P. (2024).
A modified debiased inverse-variance weighted estimator in two-sample summary-data Mendelian randomization.
{\it Statistics in Medicine} {\bfseries 43}, 5484--5496.

\bibitem[\protect\citeauthoryear{Verbanck et~al.}{2018}]{verbanck2018detection}
Verbanck, M., Chen, C.-Y., Neale, B. and Do, R. (2018).
Detection of widespread horizontal pleiotropy in causal relationships inferred from Mendelian randomization between complex traits and diseases.
{\it Nature Genetics} {\bfseries 50}, 693--698.

\bibitem[\protect\citeauthoryear{Wang and Alberding}{2024}]{wang2024powerful}
Wang, K. and Alberding, S.Y. (2024).
Powerful test of heterogeneity in two-sample summary-data Mendelian randomization.
{\it Statistics in Medicine} {\bfseries 43}, 5791--5802.

\bibitem[\protect\citeauthoryear{Xie et~al.}{2026}]{xie2026robust}
Xie, Z., Zhang, W., Wang, J. and Wu, C. (2026).
Winner’s curse free robust Mendelian randomization with summary data.
{\it Journal of the American Statistical Association}, doi:10.1080/01621459.2025.2587321.

\bibitem[\protect\citeauthoryear{Xu et~al.}{2023}]{xu2023pena}
Xu, S., Wang, P., Fung, W.K. and Liu, Z. (2023).
A novel penalized inverse-variance weighted estimator for Mendelian randomization with applications to COVID-19 outcomes.
{\it Biometrics} {\bfseries 79}, 2184--2195.

\bibitem[\protect\citeauthoryear{Xue, Shen and Pan}{Xue et~al.}{2021}]{xue2021constrain}
Xue, H., Shen, X., and Pan, W. (2021).
Constrained maximum likelihood-based Mendelian randomization robust to both correlated and uncorrelated pleiotropic effects.
{\it The American Journal of Human Genetics} {\bfseries {108}}, 1251--1269.

\bibitem[\protect\citeauthoryear{Yavorska and Burgess}{2017}]{yavorska2017Rpackage}
Yavorska, O.O., and Burgess, S. (2017).
MendelianRandomization: an R package for performing Mendelian randomization analyses using summarized data.
{\it International Journal of Epidemiology} {\bfseries 46}, 1734--1739.

\bibitem[\protect\citeauthoryear{Ye, Shao and Kang}{Ye et~al.}{2021}]{ye2021debiased}
Ye, T., Shao, J. and Kang, H. (2021).
Debiased inverse-variance weighted estimator in two-sample summary-data Mendelian randomization.
{\it The Annals of Statistics} {\bfseries 49}, 2079--2100.

\bibitem[\protect\citeauthoryear{Zhao et~al.}{2020}]{zhao2020raps}
Zhao, Q., Wang, J., Hemani, G., Bowden, J. and Small, D.S. (2020).
Statistical inference in two-sample summary-data Mendelian randomization using robust adjusted profile score.
{\it The Annals of Statistics} {\bfseries 48}, 1742--1769.
\end{thebibliography}
\end{document}


\maketitle
\newpage
To facilitate the subsequent discussions, we first introduce the notational conventions used throughout the proofs.
We use $\xrightarrow{p}$ and $\xrightarrow{d}$ to denote convergence in probability and distribution, respectively. For two sequences of random variables $\{A_n\}$ and $\{B_n\}$, we write $A_n = o_P(B_n)$ if $A_n / B_n \xrightarrow{p} 0$; $A_n = O_P(B_n)$ if $A_n/B_n$ is stochastically bounded; $A_n = \Theta_P(B_n)$ if both $A_n = O_P(B_n)$ and $B_n = O_P(A_n)$.
The sample sizes of the exposure and outcome GWAS are denoted by $n_X$ and $n_Y$, respectively.
All
sampling-error bounds below use the common scale $\nu$ from
Assumption 2 in the main text.

\section{Brief review of the RIVW estimator}
\label{app:rivw_review}

In this appendix, we present the explicit form of the RIVW estimator and its key properties. Recall from the main text that
\[
\mathcal{S}_\lambda
=
\left\{
j:
\left|
\frac{\hat{\gamma}_j}{\sigma_{X_j}}+Z_j
\right|-\lambda>0,
\quad j=1,\ldots,p
\right\},
\]
where $\lambda>0$ is a prespecified selection threshold, and
$Z_j\sim N(0,\eta^2)$ is independent noise introduced for randomized IV selection. Note that $\eta$ is a preset tuning parameter that is assumed to be bounded away from zero and infinity. For each selected SNP, we have
\[
S_j
=
\left|
\frac{\hat{\gamma}_j}{\sigma_{X_j}}+Z_j
\right|-\lambda>0.
\]

To correct for winner's-curse bias, \cite{ma2023breaking} first constructed the initial estimator
\[
\hat{\gamma}_{j,\mathrm{ini}}
=
\hat{\gamma}_j-\frac{\sigma_{X_j}}{\eta^2}Z_j.
\]
This estimator is unbiased for $\gamma_j$ and independent of the selection event. Applying the Rao--Blackwell theorem gives the minimum-variance unbiased estimator
\begin{align*}
	\hat{\gamma}_{j,\mathrm{RB}}
	&=
	E\left(
	\hat{\gamma}_{j,\mathrm{ini}}
	\mid \hat{\gamma}_j,S_j>0
	\right) \\
	&=
	\hat{\gamma}_j
	-\frac{\sigma_{X_j}}{\eta}
	\frac{\phi(A_{j,+})-\phi(A_{j,-})}
	{1-\Phi(A_{j,+})+\Phi(A_{j,-})},
\end{align*}
where
\[
A_{j,\pm}
=
-\frac{\hat{\gamma}_j}{\sigma_{X_j}\eta}
\pm\frac{\lambda}{\eta},
\]
and $\phi(\cdot)$ and $\Phi(\cdot)$ denote the standard normal density and cumulative distribution functions, respectively. We denote its conditional variance by
\[
\sigma_{X_j,\mathrm{RB}}^2
=
\operatorname{Var}\left(
\hat{\gamma}_{j,\mathrm{RB}}
\mid S_j>0
\right),
\]
which can be estimated as
\[
\hat{\sigma}_{X_j,\mathrm{RB}}^2
=
\sigma_{X_j}^2
\left[
1
-\frac{1}{\eta^2}
\frac{
	A_{j,+}\phi(A_{j,+})-A_{j,-}\phi(A_{j,-})
}{
	1-\Phi(A_{j,+})+\Phi(A_{j,-})
}
+\frac{1}{\eta^2}
\left\{
\frac{
	\phi(A_{j,+})-\phi(A_{j,-})
}{
	1-\Phi(A_{j,+})+\Phi(A_{j,-})
}
\right\}^2
\right].
\]
This estimator accounts for the variance change due to
Rao--Blackwellization and satisfies
\[
E\left(
\hat{\sigma}_{X_j,\mathrm{RB}}^2
\mid S_j>0
\right)
=
\sigma_{X_j,\mathrm{RB}}^2.
\]

Based on the preceding procedures, the RIVW estimator is defined as
\[
\hat{\beta}_{\mathrm{R}}
=
\frac{
	\displaystyle
	\sum_{j\in\mathcal{S}_\lambda}
	\sigma_{Y_j}^{-2}
	\hat{\gamma}_{j,\mathrm{RB}}\hat{\Gamma}_j
}{
	\displaystyle
	\sum_{j\in\mathcal{S}_\lambda}
	\sigma_{Y_j}^{-2}
	\left(
	\hat{\gamma}_{j,\mathrm{RB}}^2
	-\hat{\sigma}_{X_j,\mathrm{RB}}^2
	\right)
}
=
\frac{\hat{\theta}_{1,\lambda}}
{\hat{\theta}_{2,\lambda}},
\]
where
\[
\hat{\theta}_{1,\lambda}
=
\sum_{j\in\mathcal{S}_\lambda}
\sigma_{Y_j}^{-2}
\hat{\gamma}_{j,\mathrm{RB}}\hat{\Gamma}_j
\quad\text{and}\quad
\hat{\theta}_{2,\lambda}
=
\sum_{j\in\mathcal{S}_\lambda}
\sigma_{Y_j}^{-2}
\left(
\hat{\gamma}_{j,\mathrm{RB}}^2
-\hat{\sigma}_{X_j,\mathrm{RB}}^2
\right)
\]
are respectively conditionally unbiased estimators for the following quantities
\[
\theta_{1,\lambda}
=
\sum_{j\in\mathcal{S}_\lambda}
\sigma_{Y_j}^{-2}\gamma_j\Gamma_j
\quad\text{and}\quad
\theta_{2,\lambda}
=
\sum_{j\in\mathcal{S}_\lambda}
\sigma_{Y_j}^{-2}\gamma_j^2.
\]

The RIVW estimator can simultaneously correct for measurement-error bias and winner's-curse bias. \citet{ma2023breaking} established its consistency and asymptotic normality under a diverging selection threshold, $\lambda\to\infty$, and the following conditions
\[
p_\lambda\xrightarrow{p}\infty
\quad\text{and}\quad
{\kappa_\lambda}/{\lambda^2}\xrightarrow{p}\infty.
\]
{\citet{liu2026brivw} subsequently} extended RIVW to accommodate sample structure and refined the theoretical analysis of the original RIVW estimator. These refinements eliminate the requirement that $\lambda$ diverge and establish the statistical properties of the original RIVW estimator under the weaker conditions
$$
p_\lambda\xrightarrow{p}\infty
\quad\text{and}\quad
\kappa_\lambda\sqrt{p_\lambda}\xrightarrow{p}\infty.
$$

The theoretical development of the MEI test requires the consistency of the RIVW estimator. We state this result below, following two auxiliary lemmas.
These lemmas provide the technical basis for the weaker conditions stated above and support the subsequent asymptotic analysis.
Proofs of the following lemmas and Theorem~\ref{thm} are given in \citet{liu2026brivw} for the setting without sample structure.\par

\begin{lemma}[Variance comparability]
	\label{lem:rivw_variance_comparability}
	Suppose that Assumptions~1--2 hold. Then there exist constants
	$0<c_1<c_2<\infty$ such that
	\[
	c_1\sigma_{X_j}^2
	\leq
	\sigma_{X_j,\mathrm{RB}}^2
	\leq
	c_2\sigma_{X_j}^2.
	\]
\end{lemma}

\begin{lemma}[Aggregate variance-estimation error]
	\label{lem:rivw_variance_error}
	Suppose that Assumptions~1--2 hold. Then
	\[
	\left|
	\sum_{j\in\mathcal S_\lambda}
	\left(
	\hat{\sigma}_{X_j,\mathrm{RB}}^2
	-\sigma_{X_j,\mathrm{RB}}^2
	\right)
	\right|
	=
	O_P\left(\nu^2\sqrt{p_\lambda}\right).
	\]
\end{lemma}

\begin{theorem}[Consistency of the RIVW estimator]
	\label{thm}
	Suppose that Assumptions~1--3 hold,
	$p_\lambda\xrightarrow{p}\infty$, and
	$\kappa_\lambda\sqrt{p_\lambda}\xrightarrow{p}\infty$.
	Then, under $H_0$,
	\[
	\hat\beta_{\mathrm R}-\beta
	=O_P\left(
	\frac{\sqrt{\kappa_\lambda+1}}
	{\kappa_\lambda\sqrt{p_\lambda}}
	\right)
	=o_P(1).
	\]
\end{theorem}

The preceding results also imply several additional bounds that will be useful in the subsequent asymptotic analysis. Under $H_0$ and Assumptions~1--3,
\begin{align}
	\hat\theta_{1,\lambda}-\beta\hat\theta_{2,\lambda}
	&=O_P\left(\sqrt{(\kappa_\lambda+1)p_\lambda}\right),
	\label{eq:rivw_numerator_order}\\
	\hat\theta_{2,\lambda}-\theta_{2,\lambda}
	&=O_P\left(\sqrt{(\kappa_\lambda+1)p_\lambda}\right).
	\label{eq:rivw_denominator_order}
\end{align}
Furthermore, the Rao--Blackwellization arguments used in the proofs of Lemmas~\ref{lem:rivw_variance_comparability} and~\ref{lem:rivw_variance_error} yield conditional moment bounds for the individual SNP-level estimation errors and their quadratic terms. Write
\[
u_{X_j,\mathrm{RB}}=\hat\gamma_{j,\mathrm{RB}}-\gamma_j.
\]
Specifically, for $r\geq2$,
\begin{equation}
	E\left(|u_{X_j,\mathrm{RB}}|^r\mid S_j>0\right)
	=O(\nu^r),
	\label{eq:rivw_moment_bound}
\end{equation}
and, for every fixed $r\geq1$,
\begin{align}
	E\left(|\hat\sigma_{X_j,\mathrm{RB}}^2|^r\mid S_j>0\right)
	&=O(\nu^{2r}),\notag\\
	E\left(\left|u_{X_j,\mathrm{RB}}^2
	-\hat\sigma_{X_j,\mathrm{RB}}^2\right|^r\mid S_j>0\right)
	&=O(\nu^{2r}).
	\label{eq:rivw_all_quadratic_moments}
\end{align}
All moment bounds above hold uniformly in $j$ for each fixed admissible $r$. In addition, the quadratic error $u_{X_j,\mathrm{RB}}^2-\hat\sigma_{X_j,\mathrm{RB}}^2$ has conditional mean zero given $S_j>0$.

\section{Theoretical justification for decomposition of
	\texorpdfstring{$\hat{\Lambda}_{\mathrm{R,C}}$}{the centered numerator}}
\label{AppendixB}

This appendix derives a decomposition of
$\hat{\Lambda}_{\mathrm{R,C}}$ and shows that its remainder is
negligible relative to the conditional standard deviation of the
leading term. Throughout this section, we work
under the conditions stated in Theorem~\ref{thm}, using the
notation of Appendix A.

Recall from the main text, $\hat{\Lambda}_\mathrm{R}$ can be written as
\begin{align} \label{Eq:LambdaR}
	\hat{\Lambda}_\mathrm{R} =
	\hat{\theta}_{2,\lambda}
	\sum_{j \in \mathcal{S}_\lambda} \sigma_{Y_j}^{-2} \hat{\Gamma}_j
	-
	\hat{\theta}_{1,\lambda}
	\sum_{j \in \mathcal{S}_\lambda} \sigma_{Y_j}^{-2} \hat{\gamma}_{j,\mathrm{RB}},
\end{align}
which admits the decomposition:
\begin{align*}
	\hat{\Lambda}_\mathrm{R}
	&= \hat{\theta}_{2,\lambda}
	\left( \sum_{j \in \mathcal{S}_\lambda} \sigma_{Y_j}^{-2}  \hat{\Gamma}_j - \beta \sum_{j \in \mathcal{S}_\lambda} \sigma_{Y_j}^{-2} \hat{\gamma}_{j,\mathrm{RB}} \right)
	-
	\left(\hat{\theta}_{1,\lambda} - \beta \hat{\theta}_{2,\lambda} \right)
	\sum_{j \in \mathcal{S}_\lambda} \sigma_{Y_j}^{-2} \hat{\gamma}_{j,\mathrm{RB}}.
\end{align*}
To facilitate the decomposition of $\hat{\Lambda}_\mathrm{R}$, we define the quantity $T_\lambda$ along with its estimator as
\begin{align*}
	T_\lambda = \sum_{j \in \mathcal{S}_\lambda} \sigma_{Y_j}^{-2} \gamma_j, \quad
	\hat{T}_\lambda = \sum_{j \in \mathcal{S}_\lambda} \sigma_{Y_j}^{-2} \hat{\gamma}_{j,\mathrm{RB}},
\end{align*}
and let
\begin{align*}
	\Sigma_{1,\lambda}
	=  \sum_{j \in \mathcal{S}_\lambda} \sigma_{Y_j}^{-2}  \hat{\Gamma}_j - \beta \sum_{j \in \mathcal{S}_\lambda} \sigma_{Y_j}^{-2} \hat{\gamma}_{j,\mathrm{RB}},  \quad
	{\Sigma}_{2,\lambda}
	= \hat{\theta}_{1,\lambda} - \beta \hat{\theta}_{2,\lambda}.
\end{align*}
Then, $\hat{\Lambda}_\mathrm{R}$ can be expressed as:
\begin{align*}
	\hat{\Lambda}_\mathrm{R} = \hat{\theta}_{2,\lambda} \Sigma_{1,\lambda} - \hat{T}_\lambda {\Sigma}_{2,\lambda}.
\end{align*}
Further recall from the main text that
\begin{align} \label{Eq:LambdaRC}
	\hat{\Lambda}_{\mathrm{R,C}} = \hat{\Lambda}_{\mathrm{R}} + \sum_{j \in \mathcal{S}_\lambda} \sigma_{Y_j}^{-4} \hat{\sigma}_{X_j,\mathrm{RB}}^2 \hat{\Gamma}_j,
\end{align}
which can be decomposed as
\begin{align*}
	\hat{\Lambda}_{\mathrm{R,C}} = A_\lambda + \Delta_{A_\lambda},
\end{align*}
where
\begin{align*}
	A_\lambda = & \theta_{2,\lambda} \Sigma_{1,\lambda} - T_\lambda {\Sigma}_{2,\lambda},
	\quad
	\Delta_{A_\lambda} = (\hat{\theta}_{2,\lambda} - \theta_{2,\lambda}) \Sigma_{1,\lambda}
	- (\hat{T}_\lambda - T_\lambda) \Sigma_{2,\lambda}  + \sum_{j \in \mathcal{S}_\lambda} \sigma_{Y_j}^{-4} \hat{\sigma}_{X_j,\mathrm{RB}}^2 \hat{\Gamma}_j.
\end{align*}
Under $H_0$, the conditional means of
$\hat\theta_{1,\lambda}$, $\hat\theta_{2,\lambda}$, and
$\hat T_\lambda$ are $\beta\theta_{2,\lambda}$,
$\theta_{2,\lambda}$, and $T_\lambda$, respectively. Also,
\[
\operatorname{Cov}(\hat\theta_{1,\lambda},\hat T_\lambda
 \mid\mathcal S_\lambda)
=\beta\sum_{j \in \mathcal{S}_\lambda}\sigma_{Y_j}^{-4}\sigma_{X_j,\mathrm{RB}}^2\gamma_j.
\]
Then, we have
\begin{align*}
E(\hat\Lambda_{\mathrm R}\mid\mathcal S_\lambda)
&=-\beta\sum_{j \in \mathcal{S}_\lambda} \sigma_{Y_j}^{-4}\sigma_{X_j,\mathrm{RB}}^2\gamma_j,\\
E\left(\left.\sum_{j \in \mathcal{S}_\lambda}\sigma_{Y_j}^{-4}
 \hat\sigma_{X_j,\mathrm{RB}}^2\hat\Gamma_j\right|\mathcal S_\lambda\right)
&=\beta\sum_{j \in \mathcal{S}_\lambda}\sigma_{Y_j}^{-4}\sigma_{X_j,\mathrm{RB}}^2\gamma_j.
\end{align*}
Thus, the adjusted statistic $\hat\Lambda_{\mathrm {R, C}}$ is centered. Since $\Sigma_{1,\lambda}$ and
$\Sigma_{2,\lambda}$ also have conditional mean zero,
$E(A_\lambda\mid\mathcal S_\lambda)=
E(\Delta_{A_\lambda}\mid\mathcal S_\lambda)=0$.

Note that $\Sigma_{1,\lambda}$ and $\Sigma_{2,\lambda}$ can also be expressed as summations:
\begin{align*}
	&\Sigma_{1,\lambda} = \sum_{j \in \mathcal{S}_\lambda} \sigma_{Y_j}^{-2} \xi_{j, \lambda},
	\quad
	{\xi}_{j, \lambda} = \hat{\Gamma}_j - \beta \hat{\gamma}_{j,\mathrm{RB}}, \\
	& \Sigma_{2,\lambda} = \sum_{j \in \mathcal{S}_\lambda} \sigma_{Y_j}^{-2} \omega_{j, \lambda},
	\quad
	{\omega}_{j, \lambda} = \hat{\gamma}_{j,\mathrm{RB}} \hat{\Gamma}_j - \beta \left( \hat{\gamma}_{j,\mathrm{RB}}^2 - \hat{\sigma}_{X_j,\mathrm{RB}}^2 \right).
\end{align*}
Under $H_0$, $\Sigma_{2,\lambda}=\hat\theta_{1,\lambda}-\beta\hat\theta_{2,\lambda}$.
Accordingly, \({A}_\lambda\) can be written in the same form as in the main text:
\begin{align} \label{Eq:tildeA}
	A_\lambda = \sum_{j \in \mathcal{S}_\lambda} \sigma_{Y_j}^{-2} u_j,
	\quad u_j = \xi_{j, \lambda} \theta_{2,\lambda} - \omega_{j, \lambda} T_\lambda.
\end{align}
By this decomposition, $\{u_j\}_{j\in\mathcal S_\lambda}$ are mutually
independent conditional on $\mathcal S_\lambda$ and satisfy
$E(u_j\mid\mathcal S_\lambda)=0$ for each selected SNP under $H_0$.
The following lemma establishes the negligibility of
$\Delta_{A_\lambda}$ relative to the conditional standard deviation
of $A_\lambda$.

\begin{lemma}[Negligibility of the remainder]
	\label{lem:neg}
	Suppose that Assumptions~1--3 hold,
	$p_\lambda\xrightarrow{p}\infty$, and
	$\kappa_\lambda\sqrt{p_\lambda}\xrightarrow{p}\infty$.
	Then, under $H_0$,
	\[
	\frac{\Delta_{A_\lambda}}{\sqrt{V_{A,\lambda}}}
	=O_P\left(
	\frac{1}{\sqrt{p_\lambda}}
	+\frac{1}{\kappa_\lambda\sqrt{p_\lambda}}
	\right)=o_P(1),
	\]
	where $V_{A,\lambda}
	=\operatorname{Var}(A_\lambda\mid\mathcal S_\lambda)$.
\end{lemma}

\begin{proof}
	We first derive the conditional variances of $\Sigma_{1,\lambda}$
	and $\hat T_\lambda-T_\lambda$. By independence across SNPs,
	Assumptions~2--3, and
	Lemma~\ref{lem:rivw_variance_comparability},
	\begin{align*}
		\operatorname{Var}(\Sigma_{1,\lambda}\mid\mathcal S_\lambda)
		&=\sum_{j\in\mathcal S_\lambda}\sigma_{Y_j}^{-4}
		\left(\sigma_{Y_j}^2+\tau_{\alpha}^2
		+\beta^2\sigma_{X_j,\mathrm{RB}}^2\right)
		=\Theta_P(\nu^{-2}p_\lambda),\\
		\operatorname{Var}(\hat T_\lambda-T_\lambda\mid\mathcal S_\lambda)
		&=\sum_{j\in\mathcal S_\lambda}\sigma_{Y_j}^{-4}
		\sigma_{X_j,\mathrm{RB}}^2
		=\Theta_P(\nu^{-2}p_\lambda).
	\end{align*}
	Both quantities have conditional mean zero. Hence,
	\[
	\Sigma_{1,\lambda}=O_P(\nu^{-1}\sqrt{p_\lambda}),
	\qquad
	\hat T_\lambda-T_\lambda=O_P(\nu^{-1}\sqrt{p_\lambda}).
	\]
	Moreover, Equation~\eqref{eq:rivw_numerator_order} gives
	\[
	\Sigma_{2,\lambda}
	=O_P\left(\sqrt{(\kappa_\lambda+1)p_\lambda}\right).
	\]
	Equation~\eqref{eq:rivw_denominator_order}, together with
	$\theta_{2,\lambda}=\Theta_P(\kappa_\lambda p_\lambda)$,
	yields
	\[
	\hat\theta_{2,\lambda}-\theta_{2,\lambda}
	=O_P\left(\sqrt{(\kappa_\lambda+1) p_\lambda}\right).
	\]
	These relations imply that the first two terms in
	$\Delta_{A_\lambda}$ are each
	$O_P\left(\nu^{-1}p_\lambda\sqrt{\kappa_\lambda+1}\right)$.

	For the correction term in $\hat\Lambda_{\mathrm{R,C}}$,
	Equation~\eqref{eq:rivw_all_quadratic_moments} with $r=2$ gives
	\[
	E\left[
	(\hat\sigma_{X_j,\mathrm{RB}}^2-\sigma_{X_j,\mathrm{RB}}^2)^2
	\mid S_j>0\right]=O(\nu^4).
	\]
    Together with
	Lemma~\ref{lem:rivw_variance_comparability}, this implies
	$E(\hat\sigma_{X_j,\mathrm{RB}}^4\mid S_j>0)=O(\nu^4)$.
	By conditional unbiasedness and the independence of
	the outcome estimate from the exposure estimate and selection,
	\begin{align*}
		&\left|E\left(
		\sum_{j\in\mathcal S_\lambda}
		\sigma_{Y_j}^{-4}\hat\sigma_{X_j,\mathrm{RB}}^2\hat\Gamma_j
		\Bigm|\mathcal S_\lambda\right)\right|\\
		&\quad=
		\left|\beta\sum_{j\in\mathcal S_\lambda}
		\sigma_{Y_j}^{-4}\sigma_{X_j,\mathrm{RB}}^2\gamma_j\right|
		=O_P\left(\nu^{-1}p_\lambda\sqrt{\kappa_\lambda}\right),
	\end{align*}
	where the last equality follows from the uniform variance
	bounds and the Cauchy--Schwarz inequality:
	\[
	\sum_{j\in\mathcal S_\lambda}|\gamma_j|
	\leq
	\left(p_\lambda\sum_{j\in\mathcal S_\lambda}\gamma_j^2\right)^{1/2}
	=O_P\left(\nu p_\lambda\sqrt{\kappa_\lambda}\right).
	\]
	Similarly, independence across SNPs gives
	\begin{align*}
		&\operatorname{Var}\left(
		\sum_{j\in\mathcal S_\lambda}
		\sigma_{Y_j}^{-4}\hat\sigma_{X_j,\mathrm{RB}}^2\hat\Gamma_j
		\Bigm|\mathcal S_\lambda\right)\\
		&\quad\leq
		\sum_{j\in\mathcal S_\lambda}\sigma_{Y_j}^{-8}
		E(\hat\sigma_{X_j,\mathrm{RB}}^4\mid S_j>0)
		E(\hat\Gamma_j^2\mid S_j>0)
		=O_P\left\{\nu^{-2}(\kappa_\lambda+1)p_\lambda\right\},
	\end{align*}
	where the last equality uses Assumptions~2--3 and
	\[
	E(\hat\Gamma_j^2\mid S_j>0)
	=\beta^2\gamma_j^2+\sigma_{Y_j}^2+\tau_{\alpha}^2,
	\qquad
	\sum_{j\in\mathcal S_\lambda}\gamma_j^2
	=\Theta_P(\nu^2\kappa_\lambda p_\lambda).
	\]
	Combining the conditional mean and variance yields
	\[
	\sum_{j\in\mathcal S_\lambda}
	\sigma_{Y_j}^{-4}\hat\sigma_{X_j,\mathrm{RB}}^2\hat\Gamma_j
	=O_P\left\{
	\nu^{-1}p_\lambda\sqrt{\kappa_\lambda}
	+\nu^{-1}\sqrt{(\kappa_\lambda+1)p_\lambda}
	\right\}.
	\]
	Together with the orders of the other two components, this gives
	\begin{equation}
		\Delta_{A_\lambda}
		=O_P\left(\nu^{-1}p_\lambda\sqrt{\kappa_\lambda+1}\right).
		\label{eq:mei_remainder_order}
	\end{equation}

	We next consider $V_{A,\lambda}$. By Equation~\eqref{Eq:tildeA},
	\[
	u_j
	=(\theta_{2,\lambda}-T_\lambda\hat\gamma_{j,\mathrm{RB}})
	\hat\Gamma_j
	-\beta\theta_{2,\lambda}\hat\gamma_{j,\mathrm{RB}}
	+\beta T_\lambda
	(\hat\gamma_{j,\mathrm{RB}}^2-\hat\sigma_{X_j,\mathrm{RB}}^2).
	\]
	Conditional on $\mathcal S_\lambda$ and
	$\{\hat\gamma_j:j\in\mathcal S_\lambda\}$,
	only the outcome estimates remain random in this expression.
	They are independent with variances
	$\sigma_{Y_j}^2+\tau_{\alpha}^2$.
	The law of total variance therefore gives
	\begin{align*}
		V_{A,\lambda}
		&\geq\sum_{j\in\mathcal S_\lambda}\sigma_{Y_j}^{-2}
		E\left[
		(\theta_{2,\lambda}-T_\lambda\hat\gamma_{j,\mathrm{RB}})^2
		\mid\mathcal S_\lambda\right]\\
		&=\sum_{j\in\mathcal S_\lambda}\sigma_{Y_j}^{-2}
		\left\{(\theta_{2,\lambda}-T_\lambda\gamma_j)^2
		+T_\lambda^2\sigma_{X_j,\mathrm{RB}}^2\right\}.
	\end{align*}
	Write
	$H_\lambda=\sum_{j\in\mathcal S_\lambda}
	\sigma_{Y_j}^{-2}\sigma_{X_j,\mathrm{RB}}^2$.
	Completing the square, we obtain
	\begin{align*}
		&\sum_{j\in\mathcal S_\lambda}\sigma_{Y_j}^{-2}
		(\theta_{2,\lambda}-T_\lambda\gamma_j)^2
		+T_\lambda^2H_\lambda\\
		&\quad=
		\theta_{2,\lambda}^2
		\left(
		\sum_{j\in\mathcal S_\lambda}\sigma_{Y_j}^{-2}
		-\frac{T_\lambda^2}{\theta_{2,\lambda}+H_\lambda}
		\right)
		+\frac{H_\lambda^2T_\lambda^2}{\theta_{2,\lambda}+H_\lambda}\\
		&\quad\geq
		\frac{\theta_{2,\lambda}^2H_\lambda}
		{\theta_{2,\lambda}+H_\lambda}
		\sum_{j\in\mathcal S_\lambda}\sigma_{Y_j}^{-2},
	\end{align*}
	where the inequality follows from
	$T_\lambda^2\leq\theta_{2,\lambda}
	\sum_{j\in\mathcal S_\lambda}\sigma_{Y_j}^{-2}$.
	By Assumption~2 and
	Lemma~\ref{lem:rivw_variance_comparability},
	\[
	H_\lambda=\Theta_P(p_\lambda),
	\qquad
	\sum_{j\in\mathcal S_\lambda}\sigma_{Y_j}^{-2}
	=\Theta_P(\nu^{-2}p_\lambda).
	\]
	Together with
	$\theta_{2,\lambda}=\Theta_P(\kappa_\lambda p_\lambda)$,
	we have
	\begin{equation}
		\frac{1}{\sqrt{V_{A,\lambda}}}
		=O_P\left(
		\frac{\nu\sqrt{\kappa_\lambda+1}}
		{\kappa_\lambda p_\lambda^{3/2}}
		\right).
		\label{eq:mei_variance_lower_bound}
	\end{equation}
	Combining Equations~\eqref{eq:mei_remainder_order} and
	\eqref{eq:mei_variance_lower_bound} yields
	\[
	\frac{\Delta_{A_\lambda}}{\sqrt{V_{A,\lambda}}}
	=O_P\left(\frac{\kappa_\lambda+1}
	{\kappa_\lambda\sqrt{p_\lambda}}\right)
	=o_P(1)
	\]
	under the stated conditions.
\end{proof}

\section{Proof of Theorem 1}
\label{app:variance}
\label{app:normality}

This appendix proves Theorem 1 in the main text.
We first establish the consistency of the variance estimator and then prove
the asymptotic normality of the MEI test.

\begin{lemma}[Consistency of the variance estimator]
\label{lem:variance_consistency}
Suppose that Assumptions~1--4 hold,
$p_\lambda\xrightarrow{p}\infty$, and
$\kappa_\lambda\sqrt{p_\lambda}\xrightarrow{p}\infty$.
Then, under $H_0$,
\[
\frac{\hat V_{\mathrm{R,C}}}{V_{\mathrm{R,C}}}\xrightarrow{p}1.
\]
\end{lemma}

\begin{proof}
	Recall the decomposition in Appendix B,
	$\hat\Lambda_{\mathrm{R,C}}=A_\lambda+\Delta_{A_\lambda}$, where
	$A_\lambda=\sum_{j\in\mathcal S_\lambda}\sigma_{Y_j}^{-2}u_j$.
	Its conditional variance under $H_0$ is
	\[
	V_{\mathrm{R,C}}
	=\operatorname{Var}(A_\lambda\mid\mathcal S_\lambda)
	=\sum_{j\in\mathcal S_\lambda}\sigma_{Y_j}^{-4}
	\operatorname{Var}(u_j\mid\mathcal S_\lambda).
	\]
	Since $\{u_j\}_{j\in\mathcal S_\lambda}$ are mutually independent
	conditional on $\mathcal S_\lambda$ and have conditional mean zero,
	a moment estimator for $V_{\mathrm{R,C}}$ is given by
	\[
	\tilde V_{\mathrm{R,C}}
	=\sum_{j\in\mathcal S_\lambda}\sigma_{Y_j}^{-4}u_j^2.
	\]
	We first establish its consistency. For the moment calculations, write
	\[
	d_{j,\lambda}=\theta_{2,\lambda}-\gamma_jT_\lambda,
	\qquad
	V_{j,\lambda}=\sigma_{Y_j}^{-4}
	\operatorname{Var}(u_j\mid\mathcal S_\lambda).
	\]
	Recall the errors $u_{X_j,\mathrm{RB}}=\hat\gamma_{j,\mathrm{RB}}-\gamma_j$
	and $u_{Y_j}=\hat\Gamma_j-\Gamma_j$. The definitions in Appendix B give
	\begin{align*}
		u_j
		={}&(d_{j,\lambda}-T_\lambda u_{X_j,\mathrm{RB}})
		(\alpha_j+u_{Y_j})\notag\\
		&+\beta\left\{T_\lambda
		(u_{X_j,\mathrm{RB}}^2-\hat\sigma_{X_j,\mathrm{RB}}^2)
		-d_{j,\lambda}u_{X_j,\mathrm{RB}}\right\}.
		\label{Eq:decomuj}
	\end{align*}
	Conditional on the selection event, $\alpha_j+u_{Y_j}$ has mean zero and
	variance $\tau_{\alpha}^2+\sigma_{Y_j}^2$. Assumption 3 implies
	\[
	E\left(|\alpha_j+u_{Y_j}|^3\mid\mathcal S_\lambda\right)
	\leq C\nu^3.
	\]
	Together with the moment bounds in Appendix A, repeated use of
	$|a+b|^r\leq2^{r-1}(|a|^r+|b|^r)$, $r\geq1$, gives

	\begin{equation}
		E(|u_j|^3\mid\mathcal S_\lambda)
		\leq C\left\{
		|d_{j,\lambda}|^3\nu^3
		+|T_\lambda|^3\nu^6\right\}.
		\label{eq:mei_raw_moment}
	\end{equation}

	The same calculation gives the upper bound
	\[
	\operatorname{Var}(u_j\mid\mathcal S_\lambda)
	\leq C\left\{\nu^2d_{j,\lambda}^2+\nu^4T_\lambda^2\right\}.
	\]
	For the lower bound, the law of total variance gives
	\begin{align*}
		\operatorname{Var}(u_j\mid\mathcal S_\lambda)
		&\geq\sigma_{Y_j}^2
		E\{(d_{j,\lambda}-T_\lambda u_{X_j,\mathrm{RB}})^2
		\mid\mathcal S_\lambda\}\\
		&=\sigma_{Y_j}^2
		\{d_{j,\lambda}^2+T_\lambda^2\sigma_{X_j,\mathrm{RB}}^2\}.
	\end{align*}
	Thus, by Assumption 2 and
	Lemma~\ref{lem:rivw_variance_comparability}, there exist constants
	$0<c<C<\infty$ such that
	\begin{equation}
		c\{\nu^2d_{j,\lambda}^2+\nu^4T_\lambda^2\}
		\leq\operatorname{Var}(u_j\mid\mathcal S_\lambda)
		\leq C\{\nu^2d_{j,\lambda}^2+\nu^4T_\lambda^2\}.
		\label{eq:mei_summand_variance}
	\end{equation}
	Combining this with Equation~\eqref{eq:mei_raw_moment} yields

	\begin{equation}
		E\left(|\sigma_{Y_j}^{-2}u_j|^3
		\mid\mathcal S_\lambda\right)
		\leq C V_{j,\lambda}^{3/2}.
		\label{eq:mei_summand_moment}
	\end{equation}

	Next, let $C_\lambda=\theta_{2,\lambda}-\bar\gamma_\lambda T_\lambda$
	and denote the ratio in Assumption 4 by
	\[
	\varepsilon_\lambda
	=\frac{\max_{j\in\mathcal S_\lambda}(\gamma_j-\bar\gamma_\lambda)^2}
	{\sum_{j\in\mathcal S_\lambda}(\gamma_j-\bar\gamma_\lambda)^2}
	\xrightarrow{p}0.
	\]
	Since $d_{j,\lambda}=C_\lambda-T_\lambda
	(\gamma_j-\bar\gamma_\lambda)$, we have
	\begin{align*}
		\sum_{j\in\mathcal S_\lambda}d_{j,\lambda}^2
		&=p_\lambda C_\lambda^2
		+T_\lambda^2\sum_{j\in\mathcal S_\lambda}
		(\gamma_j-\bar\gamma_\lambda)^2,\\
		\max_{j\in\mathcal S_\lambda}d_{j,\lambda}^2
		&\leq2C_\lambda^2+2T_\lambda^2
		\max_{j\in\mathcal S_\lambda}(\gamma_j-\bar\gamma_\lambda)^2.
	\end{align*}
	By Equation~\eqref{eq:mei_summand_variance} and Assumptions 2 and 4,
	\begin{equation*}
		\max_{j \in \mathcal{S}_\lambda} \frac{V_{j,\lambda}}{V_{\mathrm{R,C}}}
		\leq C\frac{\max_{j \in \mathcal{S}_\lambda} d_{j,\lambda}^2+\nu^2T_\lambda^2}
		{\sum_{j \in \mathcal{S}_\lambda} d_{j,\lambda}^2+p_\lambda\nu^2T_\lambda^2}
		\leq C\left(p_\lambda^{-1}+\varepsilon_\lambda\right)=o_P(1).
		\label{eq:mei_max_variance_share}
	\end{equation*}
	It follows that

	\begin{equation}
		\frac{\sum_{j \in \mathcal{S}_\lambda} E(|\sigma_{Y_j}^{-2}u_j|^3
			\mid\mathcal S_\lambda)}{V_{\mathrm{R,C}}^{3/2}}
		\leq C\left(\max_{j \in \mathcal{S}_\lambda} \frac{V_{j,\lambda}}{V_{\mathrm{R,C}}}
		\right)^{1/2}
		\xrightarrow{p}0.
		\label{eq:mei_lyapunov_sum}
	\end{equation}

	By conditional independence and the law of large numbers for
	triangular arrays, Equation~\eqref{eq:mei_lyapunov_sum} implies
	\begin{equation}
		\frac{\tilde V_{\mathrm{R,C}}}{V_{\mathrm{R,C}}}
		\xrightarrow{p}1.
		\label{eq:mei_oracle_consistency}
	\end{equation}

	We next replace the unknown components sequentially, ultimately
	yielding a consistent estimator for $V_{\mathrm{R,C}}$.
	For the following bounds, write
	\[
	r_\lambda=\frac{\kappa_\lambda+1}
	{\kappa_\lambda\sqrt{p_\lambda}}
	=\frac1{\sqrt{p_\lambda}}
	+\frac1{\kappa_\lambda\sqrt{p_\lambda}}=o_P(1).
	\]
	By the results in Appendix A and Markov's inequality, we have
	\begin{align}
		\hat T_\lambda-T_\lambda
		&=O_P(\nu^{-1}\sqrt{p_\lambda}),&
		\hat\theta_{2,\lambda}-\theta_{2,\lambda}
		&=O_P\{\sqrt{(\kappa_\lambda+1)p_\lambda}\},\notag\\
		\sum_{j \in \mathcal{S}_\lambda}\sigma_{Y_j}^{-4}\xi_{j,\lambda}^2
		&=O_P(\nu^{-2}p_\lambda),&
		\sum_{j \in \mathcal{S}_\lambda}\sigma_{Y_j}^{-4}\omega_{j,\lambda}^2
		&=O_P\{(\kappa_\lambda+1)p_\lambda\}.
		\label{eq:mei_plugin_basic}
	\end{align}
	The lower bound \eqref{eq:mei_variance_lower_bound} implies
	\begin{equation}
		\frac1{V_{\mathrm{R,C}}}
		=O_P\left\{\frac{\nu^2(\kappa_\lambda+1)}
		{\kappa_\lambda^2p_\lambda^3}\right\}.
		\label{eq:mei_inverse_variance}
	\end{equation}
	\medskip\noindent\textbf{Step 1}: Replace $T_\lambda$ with
	$\hat T_\lambda$, yielding
	\[
	u_j^\dagger=\xi_{j,\lambda}\theta_{2,\lambda}
	-\omega_{j,\lambda}\hat T_\lambda,
	\qquad
	\tilde V_{\mathrm{R,C}}^\dagger
	=\sum_{j \in \mathcal{S}_\lambda}\sigma_{Y_j}^{-4}(u_j^\dagger)^2.
	\]
	A straightforward algebraic expansion gives
	\begin{align*}
		\tilde V_{\mathrm{R,C}}^\dagger
		={}&\tilde V_{\mathrm{R,C}}
		+\underbrace{(\hat T_\lambda-T_\lambda)^2
			\sum_{j \in \mathcal{S}_\lambda}\sigma_{Y_j}^{-4}\omega_{j,\lambda}^2}_{\text{(I)}}
		\notag\\
		&-2\underbrace{(\hat T_\lambda-T_\lambda)
			\sum_{j \in \mathcal{S}_\lambda}\sigma_{Y_j}^{-4}\omega_{j,\lambda}u_j}_{\text{(II)}}.
		\label{eq:mei_step1_expansion}
	\end{align*}
	By Equation~\eqref{eq:mei_plugin_basic},
	\[
	\text{(I)}=O_P\{\nu^{-2}(\kappa_\lambda+1)p_\lambda^2\}.
	\]
	Combining Equations~\eqref{eq:mei_oracle_consistency} and
	\eqref{eq:mei_inverse_variance}, we obtain
	\[
	\frac{\text{(I)}}{\tilde V_{\mathrm{R,C}}}
	=O_P\left\{\frac{(\kappa_\lambda+1)^2}
	{\kappa_\lambda^2p_\lambda}\right\}
	=O_P(r_\lambda^2)=o_P(1).
	\]
	For the second term, the Cauchy--Schwarz inequality gives
	\[
	\frac{|\text{(II)}|}{\tilde V_{\mathrm{R,C}}}
	\leq\left\{\frac{\text{(I)}}{\tilde V_{\mathrm{R,C}}}
	\right\}^{1/2}
	=o_P(1).
	\]
	Thus both terms are of lower asymptotic order relative to
	$\tilde V_{\mathrm{R,C}}$, and
	$\tilde V_{\mathrm{R,C}}^\dagger/
	\tilde V_{\mathrm{R,C}}\xrightarrow{p}1$.

	\medskip\noindent\textbf{Step 2}: Replace $\gamma_j^2$ by
	$\hat\gamma_{j,\mathrm{RB}}^2-\hat\sigma_{X_j,\mathrm{RB}}^2$
	(equivalently, $\theta_{2,\lambda}$ by
	$\hat\theta_{2,\lambda}$), yielding
	\[
	u_j^\ddagger=\xi_{j,\lambda}\hat\theta_{2,\lambda}
	-\omega_{j,\lambda}\hat T_\lambda,
	\qquad
	\tilde V_{\mathrm{R,C}}^\ddagger
	=\sum_{j \in \mathcal{S}_\lambda}\sigma_{Y_j}^{-4}(u_j^\ddagger)^2.
	\]
	After straightforward algebra, we have
	\begin{align*}
		\tilde V_{\mathrm{R,C}}^\ddagger
		={}&\tilde V_{\mathrm{R,C}}^\dagger
		+(\hat\theta_{2,\lambda}-\theta_{2,\lambda})^2
		\sum_{j \in \mathcal{S}_\lambda}\sigma_{Y_j}^{-4}\xi_{j,\lambda}^2\notag\\
		&+2(\hat\theta_{2,\lambda}-\theta_{2,\lambda})
		\sum_{j \in \mathcal{S}_\lambda}\sigma_{Y_j}^{-4}\xi_{j,\lambda}u_j^\dagger.
		\label{eq:mei_step2_expansion}
	\end{align*}
	By Equations~\eqref{eq:mei_plugin_basic} and \eqref{eq:mei_inverse_variance}
	and Step 1,
	\[
	\frac{(\hat\theta_{2,\lambda}-\theta_{2,\lambda})^2
		\sum_{j \in \mathcal{S}_\lambda}\sigma_{Y_j}^{-4}\xi_{j,\lambda}^2}
	{\tilde V_{\mathrm{R,C}}^\dagger}
	=O_P(r_\lambda^2)=o_P(1).
	\]
	Similarly to Step 1, the absolute value of the third term divided
	by $\tilde V_{\mathrm{R,C}}^\dagger$ is $O_P(r_\lambda)=o_P(1)$
	by the Cauchy--Schwarz inequality. It follows that
	$\tilde V_{\mathrm{R,C}}^\ddagger/
	\tilde V_{\mathrm{R,C}}^\dagger\xrightarrow{p}1$.

	\medskip\noindent\textbf{Step 3}: Replace $\beta$ by $\hat\beta_{\mathrm R}$ in $\xi_{j,\lambda}$ and $\omega_{j,\lambda}$, yielding
	\begin{align}
		\hat V_{\mathrm{R,C}}
		={}&\tilde V_{\mathrm{R,C}}^\ddagger
		+(\hat\beta_{\mathrm R}-\beta)^2
		\sum_{j\in\mathcal S_\lambda}\sigma_{Y_j}^{-4}
		\left[(\hat\gamma_{j,\mathrm{RB}}^2-\hat\sigma_{X_j,\mathrm{RB}}^2)\hat T_\lambda
		-\hat\gamma_{j,\mathrm{RB}}\hat\theta_{2,\lambda}\right]^2\notag\\
		&+2(\hat\beta_{\mathrm R}-\beta)
		\sum_{j\in\mathcal S_\lambda}\sigma_{Y_j}^{-4}
		\left[(\hat\gamma_{j,\mathrm{RB}}^2-\hat\sigma_{X_j,\mathrm{RB}}^2)\hat T_\lambda
		-\hat\gamma_{j,\mathrm{RB}}\hat\theta_{2,\lambda}\right]u_j^\ddagger.
		\label{eq:mei_step3_expansion}
	\end{align}
	To bound the second term, write
	\begin{align*}
		G_{j,\lambda}&=T_\lambda(\hat\gamma_{j,\mathrm{RB}}^2-\hat\sigma_{X_j,\mathrm{RB}}^2)-\theta_{2,\lambda}\hat\gamma_{j,\mathrm{RB}},\\
		\hat G_{j,\lambda}&=\hat T_\lambda(\hat\gamma_{j,\mathrm{RB}}^2-\hat\sigma_{X_j,\mathrm{RB}}^2)-\hat\theta_{2,\lambda}\hat\gamma_{j,\mathrm{RB}}.
	\end{align*}
	Then the replacement error is
	\begin{equation*}
		\hat u_j-u_j^\ddagger=(\hat\beta_{\mathrm R}-\beta)\hat G_{j,\lambda}.
		\label{eq:mei_beta_replacement}
	\end{equation*}
	We first bound the term involving $G_{j,\lambda}$. Let $d_{j,\lambda}=\theta_{2,\lambda}-\gamma_jT_\lambda$. Expanding in $u_{X_j,\mathrm{RB}}$ gives
	\begin{align*}
		G_{j,\lambda}
		={}&-\gamma_jd_{j,\lambda}+(T_\lambda\gamma_j-d_{j,\lambda})u_{X_j,\mathrm{RB}}\\
		&+T_\lambda(u_{X_j,\mathrm{RB}}^2-\hat\sigma_{X_j,\mathrm{RB}}^2).
	\end{align*}
	For the first term, expanding about the weighted mean $T_\lambda/\sum_{j\in\mathcal S_\lambda}\sigma_{Y_j}^{-2}$ and using Equation~\eqref{eq:mei_summand_variance} yields
	\[
	\frac{\sum_{j\in\mathcal S_\lambda}\sigma_{Y_j}^{-4}\gamma_j^2d_{j,\lambda}^2}{V_{\mathrm{R,C}}}
	\leq C\left\{\kappa_\lambda+\min(1,\kappa_\lambda)\frac{\max_{j\in\mathcal S_\lambda}(\gamma_j-\bar\gamma_\lambda)^2}{\nu^2}\right\}.
	\]
	Applying the moment bounds in Appendix A to the remaining two terms, we obtain
	\begin{equation*}
		\frac{\sum_{j\in\mathcal S_\lambda}\sigma_{Y_j}^{-4}G_{j,\lambda}^2}{V_{\mathrm{R,C}}}
		=O_P\left\{1+\kappa_\lambda+\min(1,\kappa_\lambda)\frac{\max_{j\in\mathcal S_\lambda}(\gamma_j-\bar\gamma_\lambda)^2}{\nu^2}\right\}.
		\label{eq:mei_multiplier_bound}
	\end{equation*}
	Since $\max_{j\in\mathcal S_\lambda}(\gamma_j-\bar\gamma_\lambda)^2=O_P(\varepsilon_\lambda\nu^2\kappa_\lambda p_\lambda)$ and Appendix A gives $(\hat\beta_{\mathrm R}-\beta)^2=O_P\{(\kappa_\lambda+1)/(\kappa_\lambda^2p_\lambda)\}$, we obtain
	\begin{equation*}
		\frac{(\hat\beta_{\mathrm R}-\beta)^2\sum_{j\in\mathcal S_\lambda}\sigma_{Y_j}^{-4}G_{j,\lambda}^2}{V_{\mathrm{R,C}}}
		=O_P(r_\lambda^2+\varepsilon_\lambda)=o_P(1).
		\label{eq:mei_true_multiplier_error}
	\end{equation*}
	Next, the difference between $\hat G_{j,\lambda}$ and $G_{j,\lambda}$ is
	\begin{align*}
		\hat G_{j,\lambda}-G_{j,\lambda}
		=(\hat T_\lambda-T_\lambda)(\hat\gamma_{j,\mathrm{RB}}^2-\hat\sigma_{X_j,\mathrm{RB}}^2)
		-(\hat\theta_{2,\lambda}-\theta_{2,\lambda})\hat\gamma_{j,\mathrm{RB}}.
	\end{align*}
	Expanding $\gamma_j$ about $\bar\gamma_\lambda$ gives $\sum_{j\in\mathcal S_\lambda}\gamma_j^4=O_P\{\nu^4\kappa_\lambda^2(p_\lambda+\varepsilon_\lambda p_\lambda^2)\}$. Together with the moment bounds in Appendix A and Equations~\eqref{eq:mei_plugin_basic} and \eqref{eq:mei_inverse_variance}, this implies
	\begin{align*}
		\frac{(\hat\beta_{\mathrm R}-\beta)^2\sum_{j\in\mathcal S_\lambda}\sigma_{Y_j}^{-4}(\hat G_{j,\lambda}-G_{j,\lambda})^2}{V_{\mathrm{R,C}}}
		=O_P\{r_\lambda^2(p_\lambda^{-1}+\varepsilon_\lambda+r_\lambda^2)\}=o_P(1).
		\label{eq:mei_estimated_multiplier_error}
	\end{align*}
	Combining these results and Step 2, we obtain
	\[
	\frac{(\hat\beta_{\mathrm R}-\beta)^2\sum_{j\in\mathcal S_\lambda}\sigma_{Y_j}^{-4}\hat G_{j,\lambda}^2}{\tilde V_{\mathrm{R,C}}^\ddagger}=O_P(r_\lambda^2+\varepsilon_\lambda)=o_P(1).
	\]
	For the cross term, the Cauchy--Schwarz inequality gives
	\[
	\frac{\left|(\hat\beta_{\mathrm R}-\beta)\sum_{j\in\mathcal S_\lambda}\sigma_{Y_j}^{-4}\hat G_{j,\lambda}u_j^\ddagger\right|}{\tilde V_{\mathrm{R,C}}^\ddagger}
	\leq\left\{\frac{(\hat\beta_{\mathrm R}-\beta)^2\sum_{j\in\mathcal S_\lambda}\sigma_{Y_j}^{-4}\hat G_{j,\lambda}^2}{\tilde V_{\mathrm{R,C}}^\ddagger}\right\}^{1/2}=o_P(1).
	\]
	Thus the additional terms in Equation~\eqref{eq:mei_step3_expansion} are negligible relative to $\tilde V_{\mathrm{R,C}}^\ddagger$, giving $\hat V_{\mathrm{R,C}}/\tilde V_{\mathrm{R,C}}^\ddagger\xrightarrow{p}1$.

	Combining the three steps and Equation~\eqref{eq:mei_oracle_consistency},
	we conclude that
	\[
	\frac{\hat V_{\mathrm{R,C}}}{V_{\mathrm{R,C}}}\xrightarrow{p}1,
	\]
	under Assumptions 1--4, $p_\lambda\xrightarrow{p}\infty$, and
	$\kappa_\lambda\sqrt{p_\lambda}\xrightarrow{p}\infty$.
	The bounds for the quadratic terms in the three steps also give
	\begin{equation*}
		\frac{\sum_{j\in\mathcal S_\lambda}\sigma_{Y_j}^{-4}(\hat u_j-u_j)^2}
		{V_{\mathrm{R,C}}}\xrightarrow{p}0.
		\label{eq:mei_full_plugin_error}
	\end{equation*}
\end{proof}

\begin{proof}[Proof of Theorem 1]
	We first consider $A_\lambda$. Under $H_0$, conditional on
	$\mathcal S_\lambda$, the variables $X_j=\sigma_{Y_j}^{-2}u_j$
	are independent and have mean zero. For every $\epsilon>0$,

	\begin{align*}
		&\frac{1}{V_{\mathrm{R,C}}}
		\sum_{j\in\mathcal S_\lambda}
		E\left\{
		X_j^2 I\left(|X_j|>\epsilon\sqrt{V_{\mathrm{R,C}}}\right)
		\mid\mathcal S_\lambda
		\right\}\\
		&\qquad\leq
		\frac{\epsilon^{-1}}{V_{\mathrm{R,C}}^{3/2}}
		\sum_{j\in\mathcal S_\lambda}
		E\left(|X_j|^3\mid\mathcal S_\lambda\right)
		=o_P(1)
	\end{align*}

	by Equation~\eqref{eq:mei_lyapunov_sum}. Hence the conditional Lindeberg
	condition holds, and the Lindeberg--Feller theorem gives
	\[
	\frac{A_\lambda}{\sqrt{V_{\mathrm{R,C}}}}
	\xrightarrow{d}N(0,1).
	\]
	By Lemma~\ref{lem:neg} in Appendix B and Lemma~\ref{lem:variance_consistency},
	\[
	\frac{\Delta_{A_\lambda}}{\sqrt{V_{\mathrm{R,C}}}}=o_P(1),
	\qquad
	\frac{\hat V_{\mathrm{R,C}}}{V_{\mathrm{R,C}}}
	\xrightarrow{p}1.
	\]
	Combining these results with
	$\hat\Lambda_{\mathrm{R,C}}=A_\lambda+\Delta_{A_\lambda}$
	and applying Slutsky's theorem, we obtain
	\[
	Z_{\mathrm{ME}}
	=
	\left(
	\frac{A_\lambda}{\sqrt{V_{\mathrm{R,C}}}}
	+
	\frac{\Delta_{A_\lambda}}{\sqrt{V_{\mathrm{R,C}}}}
	\right)
	\sqrt{
	\frac{V_{\mathrm{R,C}}}{\hat V_{\mathrm{R,C}}}
	}
	\xrightarrow{d}N(0,1).
	\]
\end{proof}
\section{Conditional expectation of \texorpdfstring{$\hat\Lambda_{\mathrm{R,C}}$}{Lambda R,C} under \texorpdfstring{$H_1$}{H1}}
\label{app:coding_expectation}

In this appendix, we calculate the conditional expectation of $\hat\Lambda_{\mathrm{R,C}}$ under $H_1$. Combining Equations~\eqref{Eq:LambdaR} and \eqref{Eq:LambdaRC}, we write
\begin{align*}
	\hat\Lambda_{\mathrm{R,C}}
	={}&\hat\theta_{2,\lambda}\sum_{j\in\mathcal S_\lambda}\sigma_{Y_j}^{-2}\hat\Gamma_j
	-\hat\theta_{1,\lambda}\hat T_\lambda
	+\sum_{j\in\mathcal S_\lambda}\sigma_{Y_j}^{-4}\hat\sigma_{X_j,\mathrm{RB}}^2\hat\Gamma_j.
\end{align*}
Conditional on $\mathcal S_\lambda$ and the true effects $\bm\gamma,\bm\alpha$, the exposure and outcome sampling errors are independent. By the conditional unbiasedness results in Appendix A, we have
\begin{align*}
	E(\hat\theta_{1,\lambda}\mid\mathcal S_\lambda,\bm\gamma,\bm\alpha)
	&=\sum_{j\in\mathcal S_\lambda}\sigma_{Y_j}^{-2}\gamma_j\Gamma_j,\\
	E(\hat\theta_{2,\lambda}\mid\mathcal S_\lambda,\bm\gamma,\bm\alpha)
	&=\theta_{2,\lambda},\\
	E(\hat T_\lambda\mid\mathcal S_\lambda,\bm\gamma,\bm\alpha)
	&=T_\lambda,\\
	\operatorname{Cov}(\hat\theta_{1,\lambda},\hat T_\lambda\mid\mathcal S_\lambda,\bm\gamma,\bm\alpha)
	&=\sum_{j\in\mathcal S_\lambda}\sigma_{Y_j}^{-4}\sigma_{X_j,\mathrm{RB}}^2\Gamma_j.
\end{align*}
The covariance term is canceled by the expectation of the correction term in $\hat\Lambda_{\mathrm{R,C}}$. Therefore,
\begin{align*}
	E(\hat\Lambda_{\mathrm{R,C}}\mid\mathcal S_\lambda,\bm\gamma,\bm\alpha)
	={}&\theta_{2,\lambda}\sum_{j\in\mathcal S_\lambda}\sigma_{Y_j}^{-2}\Gamma_j
	-T_\lambda\sum_{j\in\mathcal S_\lambda}\sigma_{Y_j}^{-2}\gamma_j\Gamma_j.
\end{align*}
Substituting $\Gamma_j=\beta\gamma_j+\alpha_j$ and canceling the diagonal terms gives
\begin{align*}
	E(\hat\Lambda_{\mathrm{R,C}}\mid\mathcal S_\lambda,\bm\gamma,\bm\alpha)
	={}&\theta_{2,\lambda}\sum_{j\in\mathcal S_\lambda}\sigma_{Y_j}^{-2}\alpha_j
	-T_\lambda\sum_{j\in\mathcal S_\lambda}\sigma_{Y_j}^{-2}\gamma_j\alpha_j\notag\\
	={}&\sum_{\substack{j,k\in\mathcal S_\lambda\\j\ne k}}
	\sigma_{Y_j}^{-2}\sigma_{Y_k}^{-2}
	(\gamma_j^2\alpha_k-\gamma_j\alpha_j\gamma_k).
\end{align*}

We now take expectation with respect to the true effects. Suppose that the pairs $(\gamma_j,\alpha_j)$ are independent across SNPs and have finite second moments, with the standard errors treated as fixed. Since selection is performed separately for each SNP, the pairs remain independent conditional on $\mathcal S_\lambda$. Taking expectation in the preceding equation, we obtain
\begin{align}
	E(\hat\Lambda_{\mathrm{R,C}}\mid\mathcal S_\lambda)
	=\sum_{\substack{j,k\in\mathcal S_\lambda\\j\ne k}}
	\sigma_{Y_j}^{-2}\sigma_{Y_k}^{-2}
	\big[&E(\gamma_j^2\mid S_j>0)E(\alpha_k\mid S_k>0)\notag\\
	&-E(\gamma_j\alpha_j\mid S_j>0)E(\gamma_k\mid S_k>0)\big].
	\label{eq:coding_general_expectation}
\end{align}
Following the notation in the main text, we denote the
conditional moments of the selected SNPs by
\begin{align*}
	E(\gamma_j\mid S_j>0)&=\mu_{\gamma\mid\mathcal S_\lambda}, \quad
	E(\alpha_j\mid S_j>0)=\mu_{\alpha\mid\mathcal S_\lambda},\\
	\operatorname{Var}(\gamma_j\mid S_j>0)&=\tau_{\gamma\mid\mathcal S_\lambda}^2, \quad
	\operatorname{Var}(\alpha_j\mid S_j>0)=\tau_{\alpha\mid\mathcal S_\lambda}^2.
\end{align*}
Substituting these moments into Equation~\eqref{eq:coding_general_expectation} and using
\[
\sum_{\substack{j,k\in\mathcal S_\lambda\\j\ne k}}\sigma_{Y_j}^{-2}\sigma_{Y_k}^{-2}
=\left(\sum_{j\in\mathcal S_\lambda}\sigma_{Y_j}^{-2}\right)^2
-\sum_{j\in\mathcal S_\lambda}\sigma_{Y_j}^{-4},
\]
we obtain
\begin{align*}
	E(\hat\Lambda_{\mathrm{R,C}}\mid\mathcal S_\lambda)
	=\left[\left(\sum_{j\in\mathcal S_\lambda}\sigma_{Y_j}^{-2}\right)^2
	-\sum_{j\in\mathcal S_\lambda}\sigma_{Y_j}^{-4}\right]
	\left(
	\mu_{\alpha\mid\mathcal S_\lambda}\tau_{\gamma\mid\mathcal S_\lambda}^2
	-\rho_{\gamma,\alpha\mid\mathcal S_\lambda}
	\mu_{\gamma\mid\mathcal S_\lambda}
	\tau_{\alpha\mid\mathcal S_\lambda}
	\tau_{\gamma\mid\mathcal S_\lambda}\right).
	\label{eq:coding_common_expectation}
\end{align*}

\setcounter{table}{0}
\renewcommand{\tablename}{Web Table}
\renewcommand{\thetable}{S\arabic{table}}

\clearpage
\begingroup
\captionsetup[table]{labelfont=bf}
\begin{table}[H]

\centering
\caption{Estimated type I error rates (\%) of various tests at the nominal significance level of 0.05 under $\mu_x=0.005$ based on $10,000$ replicates.}
\label{tab:type1error_mu004}
\begin{tabular*}{\textwidth}{@{\extracolsep{\fill}}cccccccc@{}}
\toprule
$n$ & $q$ & $\pi_1$ & $Z_{\mathrm{ME}}^{\mathrm{C}}$ & $Z_{\mathrm{ME}}^{\mathrm{M}}$ & $Z_{\mathrm{ME}}^{\mathrm{P}}$ & $Z_{\mathrm{E}}$ & $Q^*$ \\
\midrule
250,000 & 0\% & 5\% & 5.11 & 5.40 & 4.90 & 12.47 & 7.87 \\
 &  & 10\% & 4.98 & 4.82 & 5.34 & 20.00 & 8.81 \\
 &  & 20\% & 5.28 & 5.26 & 5.30 & 34.71 & 11.26 \\

 & 10\% & 5\% & 5.65 & 5.33 & 5.70 & 10.22 & 80.05 \\
 &  & 10\% & 5.16 & 5.00 & 5.30 & 16.33 & 95.35 \\
 &  & 20\% & 5.03 & 5.03 & 5.05 & 26.49 & 99.87 \\

500,000 & 0\% & 5\% & 4.92 & 4.98 & 4.80 & 31.86 & 8.08 \\
 &  & 10\% & 5.25 & 4.93 & 4.91 & 56.60 & 9.51 \\
 &  & 20\% & 5.18 & 5.08 & 4.94 & 83.98 & 12.10 \\

 & 10\% & 5\% & 4.88 & 5.16 & 4.97 & 20.61 & 100.00 \\
 &  & 10\% & 5.23 & 5.39 & 5.24 & 35.53 & 100.00 \\
 &  & 20\% & 4.98 & 5.03 & 5.19 & 62.50 & 100.00 \\
\bottomrule
\end{tabular*}
\par\vspace{0.35\baselineskip}
\parbox{\textwidth}{%
	\footnotesize\justifying\noindent
	*For \(Q\), the results under \(q=10\%\) represent its empirical power to detect balanced pleiotropy under the InSIDE assumption.
	\par
}
\end{table}
\endgroup

\clearpage
\begin{figure}[H]
	\centering
	\includegraphics[width=\linewidth,height=0.76\textheight,keepaspectratio]{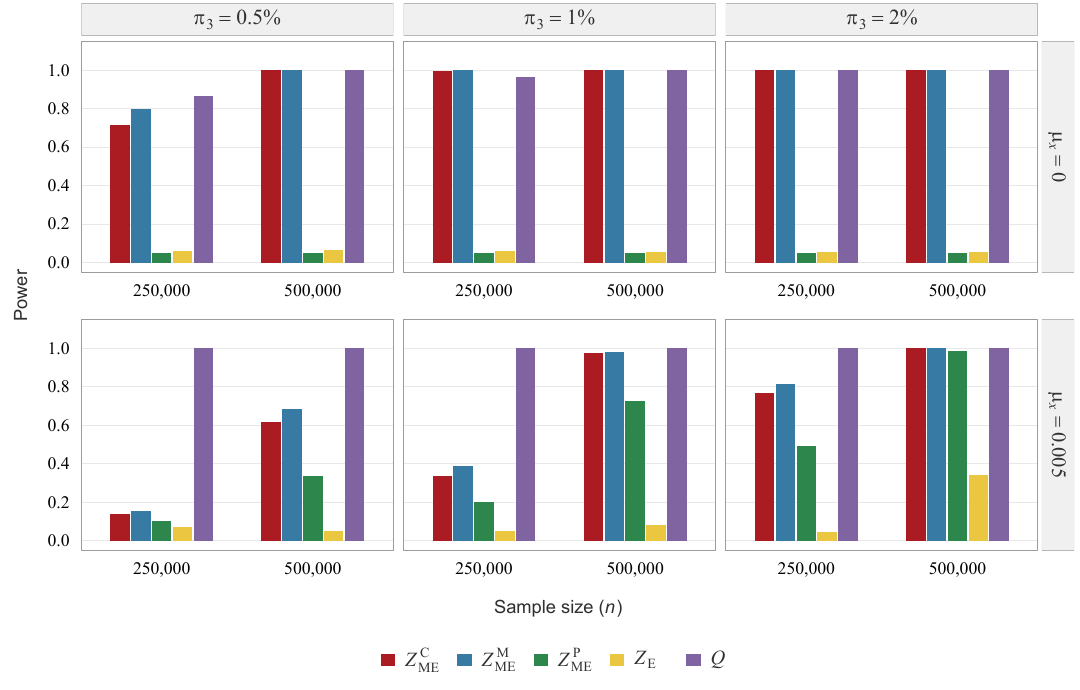}
	\caption{Estimated power of various tests to detect directional pleiotropy ($r=1$) based on $10,000$ replicates. The proportion of relevant IVs without directional or correlated pleiotropy is set at $\pi_1=10\%$. Balanced pleiotropic IVs are present ($q=10\%$).}
	\label{fig:PowerDirect_q10_pi1_10}
\par\smallskip\noindent
\figalttext{Grouped bar charts comparing the power of the five tests to detect directional pleiotropy in the presence of balanced pleiotropy, for proportions of directional pleiotropic IVs of 0.5, 1 and 2 percent, GWAS sample sizes of 250,000 and 500,000, and mean SNP--exposure effects among relevant IVs of 0 and 0.005.}
\end{figure}

\clearpage
\begin{figure}[H]
	\centering
	\includegraphics[width=\linewidth,height=0.76\textheight,keepaspectratio]{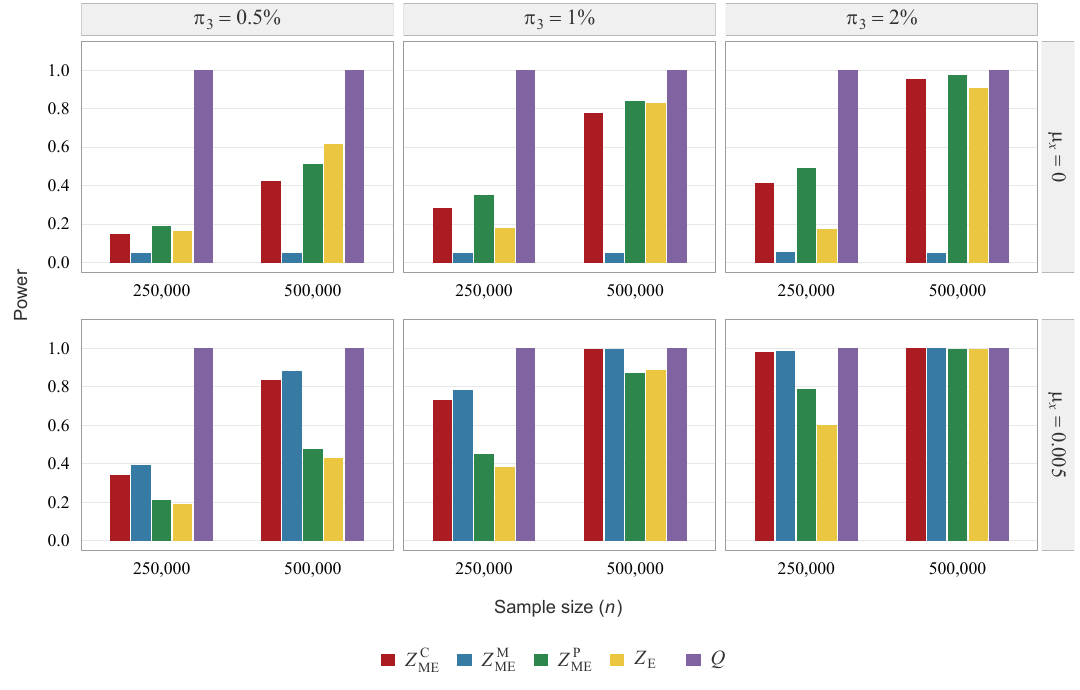}
	\caption{Estimated power of various tests to detect correlated pleiotropy ($r=0$) based on $10,000$ replicates. The proportion of relevant IVs without directional or correlated pleiotropy is set at $\pi_1=10\%$. Balanced pleiotropic IVs are present ($q=10\%$).}
	\label{fig:PowerCorr_q10_pi1_10}
\par\smallskip\noindent
\end{figure}

\clearpage
\begin{figure}[H]
	\centering
	\includegraphics[width=\linewidth,height=0.76\textheight,keepaspectratio]{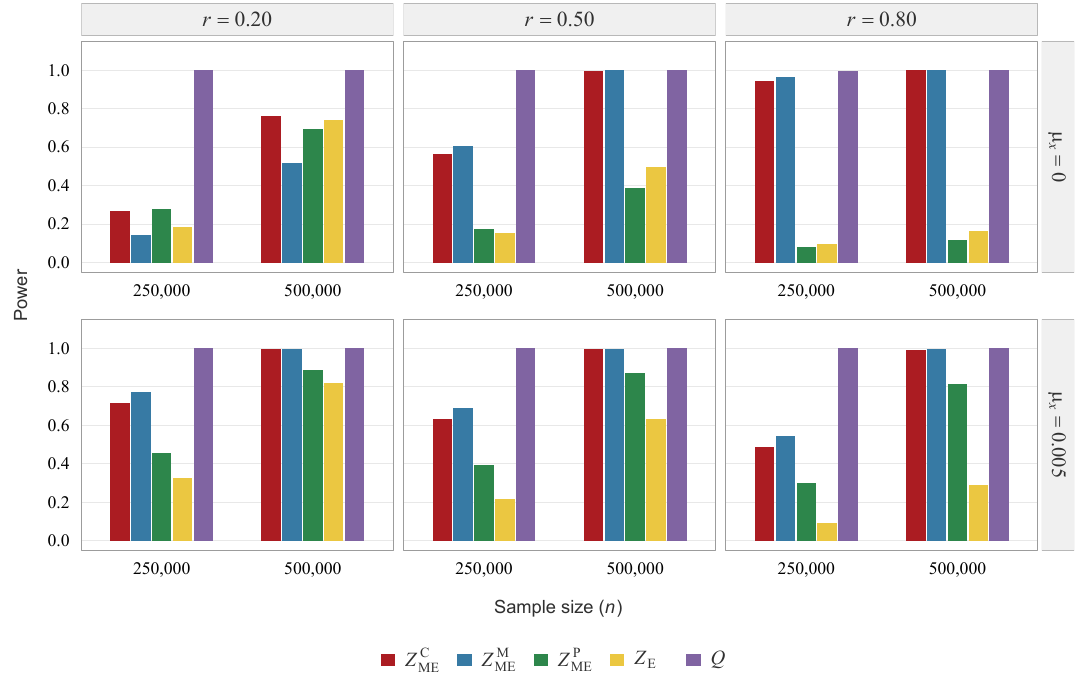}
	\caption{Estimated power of various tests under varying proportions of directional and correlated pleiotropy based on $10,000$ replicates. The proportion of relevant IVs without directional or correlated pleiotropy is set at $\pi_1=10\%$, and the total proportion of directional and correlated pleiotropic IVs is set at $\pi_3=1\%$. Balanced pleiotropic IVs are present ($q=10\%$).}
	\label{fig:PowerDC_q10_pi3_010}
\par\smallskip\noindent
\figalttext{Grouped bar charts comparing the power of the five tests in the presence of balanced pleiotropy as the proportion of directional pleiotropic IVs among all IVs with directional or correlated pleiotropy increases from 20 to 50 to 80 percent, for GWAS sample sizes of 250,000 and 500,000, and mean SNP--exposure effects among relevant IVs of 0 and 0.005.}
\end{figure}